\pdfoutput=1
\documentclass[11pt, a4paper]{article}
\usepackage[utf8]{inputenc}
\usepackage[T1]{fontenc}
\usepackage{amsmath, amssymb, amsfonts, bm, cite, float}
\usepackage{graphicx}
\usepackage{booktabs, longtable, array, calc}
\usepackage{geometry}
\usepackage[hidelinks]{hyperref}

\allowdisplaybreaks
\numberwithin{equation}{section}

\newcommand{\papertitle}{One-Loop Fluctuation Response Along a Constrained Noncommutative Modulus in the Lorentzian IIB Matrix Model}
\hypersetup{pdftitle={\papertitle},pdfauthor={Tetsuyuki Muramatsu}}

\begin{document}

\hypersetup{pageanchor=false}
\begin{titlepage}
\vspace{2cm}
\begin{center}
{\Large\bf \papertitle}
\vspace{2cm}

{\large Tetsuyuki Muramatsu\footnote{The views expressed in this paper are solely those of the author and do not necessarily reflect those of the affiliated organization.}} \\

\vspace{1cm}
{\it Tokai National Higher Education and Research System \\
Furo-cho, Chikusa-ku, Nagoya, 464-8601, Japan} \\
\vspace{2cm}
\end{center}

\begin{center}
{\bf Abstract}
\end{center}
We study a local constrained problem at matrix size \(N=5\) in the Lorentzian IIB (IKKT) matrix model and numerically continue a one-dimensional physical modulus. For an exact fixed-\(Q_\rho\) constrained solution family, the stationary identity implies constant classical action; the 23 accepted backgrounds realize this relation to the quoted precision. Across the sampled branch, the spectrum of the Lorentzian extent operator remains common to numerical precision while microscopic noncommutative invariants vary.

We evaluate finite-dimensional one-loop fluctuations along this branch. At fixed bosonic background, the fermionic Grassmann integral is exact and gives \(\operatorname{Pf}\mathcal M_F\); its magnitude is invariant under color-gauge and connected proper-Lorentz transformations and varies along the modulus. The complete boson--ghost--fermion quotient measure defines a Lorentz-representative-independent density one-form rather than a scalar potential. Thus quantum fluctuation data distinguish backgrounds unresolved by the coarse quadratic spectrum.

We also apply a framed higher-\(L\) SO(5) structural probe. At the reference background, the ordered \(3+6\) spectral separation is larger at \(L=2,3\) than at \(L=1\), and larger still in the canonical large-\(L\) limit. This is a frame-defined diagnostic, not an intrinsic observable or a higher-\(N\) solution. The Pfaffian and higher-\(L\) responses are closely correlated along the sampled branch, which we treat only as descriptive. These results suggest that microscopic noncommutative structure and quantum fluctuations may contain information relevant to dimensional dynamics, without establishing dimensional selection.

\end{titlepage}
\hypersetup{pageanchor=true}

\newpage
\section{Introduction}\label{introduction}

The IIB matrix model was proposed as a nonperturbative formulation of superstring theory and has been studied as a framework in which spacetime may emerge dynamically from matrix degrees of freedom \cite{Ishibashi:1996xs}. A central question is how a lower-dimensional spacetime may be selected from the ten-dimensional matrices. Lorentzian numerical studies have used the eigenvalues of the spatial moment-of-inertia tensor to characterize the extent of space and have reported behavior associated with an \(SO(9)\to SO(3)\) pattern \cite{Kim:2011cr,Anagnostopoulos:2023,Nishimura:2019}. Concrete finite-matrix classical solutions, including expanding configurations with noncommutative extra-dimensional structure, have also been studied \cite{Hatakeyama:2020}, and recent complex Langevin work has reported a smooth real \((3+1)\)-dimensional expanding phase in a deformed Lorentzian model \cite{Anagnostopoulos:2026}.

In this paper, we consider the coarse quadratic geometry of a ten-dimensional Lorentzian background described by

\begin{equation}
G_{\mu\nu}
=
\frac{1}{N}\operatorname{Tr}(A_\mu A_\nu).
\end{equation}

By raising one index with the Lorentz metric, we define the \textbf{Lorentzian extent operator}

\begin{equation}
K^\mu{}_{\nu}
=
\eta^{\mu\rho}G_{\rho\nu}.
\end{equation}

Its eigenvalue multiset, which we call the \(K\)-spectrum, gives Lorentz-frame-independent quadratic spectral data. These data need not contain all microscopic information in the matrix configuration. We therefore ask whether different microscopic noncommutative structures can lie behind the same coarse quadratic spectral data, and whether quantum fluctuations can distinguish them.

To address this question in a finite-dimensional setting, we study the Lorentzian IIB matrix model at matrix size \(N=5\), with the decoupled \(U(1)\) center-of-mass sector removed. At this matrix size, the interacting matrix content decomposes into two irreducible sectors with respect to an SO(5) structural frame, while the constrained classical problem and one-loop fluctuation operators remain directly finite dimensional. SO(5)-covariant finite matrix geometries and fuzzy four-sphere constructions are well established \cite{Ramgoolam:2001,Kimura:2002,Steinacker:2015,Sperling:2017}, and finite-size matrix descriptions of curved spacetimes have also been developed recently \cite{HattoriSekoTsuchiya:2026}. The representation-theoretic machinery itself is therefore not the novelty pursued here. Our use of SO(5) is different from a fuzzy-\(S^4\) saddle or stabilization calculation: a fixed structural frame is used to decompose and track the microscopic content of the same constrained \(N=5\) branch.

After removing color-gauge and Lorentz-frame redundancies, we numerically continue a one-dimensional physical modulus. The exact constrained stationary-family identity implies constancy of the classical action at fixed \(Q_\rho\), and the accepted backgrounds realize this relation to the quoted numerical precision. On the 23 accepted backgrounds spanning the sampled local branch, the full \(K\)-spectrum remains common to the stated numerical precision, while higher-order and commutator-sensitive invariants vary. Thus the microscopic noncommutative structure changes continuously even though the coarse quadratic spectral data do not resolve the deformation.

Quantum effects on matrix-model moduli and background-dependent one-loop actions are known from early studies of IIB matrix-model moduli \cite{Aoki:1998}, one-loop analyses of noncommutative and cosmological backgrounds \cite{Blaschke:2011,Steinacker:2023,BattistaSteinacker:2024}, and recent weak-coupling studies of background and fluctuation scales \cite{SteinackerWeak:2026}. Continuous equal-energy matrix-background families with different fluctuation spectra are also known in related settings \cite{SperlingSteinacker:2018}, while determinant--Pfaffian moduli-space measures occur in the polarised model \cite{HartnollLiu:2025}. Quantum behavior along commuting D-instanton flat directions has also been revisited recently in the bosonic IIB matrix model \cite{Chen:2026}. Thus neither generic quantum lifting of matrix-model moduli nor background-dependent fluctuation spectra are claimed as new here.

Our question is more specific. We evaluate the fluctuations along the same local noncommutative Lorentzian branch for which the complete coarse \(K\)-spectrum fails, within numerical precision, to resolve the microscopic deformation. The Majorana--Weyl Pfaffian magnitude varies along the modulus and, as shown in Appendix C, is a scalar under color-gauge transformations and connected proper-Lorentz transformations. Fermionic Pfaffians are already known to carry geometric information in the IIB matrix model \cite{NishimuraVernizzi:2000}; here the relevant quantity is the branchwise variation of the Pfaffian magnitude on the constrained physical modulus. The complete boson--ghost--fermion quotient measure defines a density one-form on the solution curve, whereas the combined hard-determinant expression in a chosen auxiliary Lorentz section is retained only as a fixed-section diagnostic.

The noncompact Lorentz orbit of the full Lorentzian IIB matrix integral has recently been treated by an explicit Faddeev--Popov prescription \cite{Asano:2025}. Our local quotient instead removes Lorentz-frame tangent directions when classifying the constrained stationary branch and constructing its local density. The specific contribution of this paper is therefore the coexistence, on one local finite-\(N\) Lorentzian physical branch, of a common coarse \(K\)-spectrum, continuously varying microscopic noncommutative data, and an intrinsic fermionic response. We are not aware of a previous Lorentzian IIB analysis exhibiting this combination.

We also introduce a framed higher-\(L\) structural representation probe for the same \(N=5\) backgrounds. It is defined relative to a chosen SO(5) embedding \(\iota\), transported together with the background under color-gauge transformations, and is therefore a diagnostic of the framed pair \((A,\iota)\), not an intrinsic observable of \(A\) alone. The relative quadratic-trace weight of the \(\mathbf{14}\) sector with respect to the \(\mathbf{10}\) sector is

\begin{equation}
\rho_L
=
\frac{4L(L+3)+5}{42},
\end{equation}

which increases exactly with \(L\). This makes a redistribution hidden in the \(L=1\) quadratic trace visible as a higher-\(L\) spectral response. At the reference background, the ordered \(3+6\) diagnostic---an average separation, not an exact degeneracy---is numerically larger at \(L=2\) and \(L=3\) than at \(L=1\), and larger still in the canonical large-\(L\) limit. Here \(L\) is only the representation label; the dynamical matrix size remains \(N=5\). The underlying SO(5) and fuzzy-\(S^4\) representation theory is standard \cite{Ramgoolam:2001,Kimura:2002,Sperling:2017}; the present use is to transport the fixed \(N=5\) coefficient data through this representation family as a framed structural diagnostic. The constrained branch is not obtained from the mass deformation or localization construction of the polarised model \cite{HartnollLiu:2025}.

Finally, the intrinsic Pfaffian response and the framed higher-\(L\) response show a close numerical correspondence along the sampled modulus. Because the comparison is made on a short one-dimensional branch, we treat it as descriptive rather than as evidence for a common dynamical mechanism or analytic identity. The SO(5)-resolved redistribution is already present classically, and neither response is taken to generate the other. The comparison does not establish dimensional selection. Sections 2--7 present the construction and analysis; Sections 8 and 9 discuss the implications and limitations. Technical derivations, numerical evidence, conventions, and diagnostic details are collected in Appendices A--F.

\section{\texorpdfstring{IIB matrix model and the SO(5) decomposition at \(N=5\)}{IIB matrix model and the SO(5) decomposition at N=5}}\label{iib-matrix-model-and-the-so5-decomposition-at-n5}

\subsection{IIB matrix model and coarse quadratic geometry}\label{iib-matrix-model-and-coarse-quadratic-geometry}

The IIB matrix model consists of ten \(N\times N\) Hermitian matrices

\begin{equation}
A_\mu,
\qquad
\mu=0,\ldots,9,
\end{equation}

and an \(N\times N\) matrix-valued ten-dimensional Majorana--Weyl spinor

\begin{equation}
\psi_\alpha,
\qquad
\alpha=1,\ldots,16.
\end{equation}

Here \(N\) is the matrix size, equivalently the \(N\) in the \(U(N)\) color symmetry. The \(U(1)\) trace part is the decoupled center-of-mass sector, so the interacting traceless sector is the adjoint of \(SU(N)\), with \(N^2-1\) real generators.

We use the mostly-plus Lorentz metric

\begin{equation}
\eta_{\mu\nu}
=
\operatorname{diag}(-1,+1,\ldots,+1),
\end{equation}

and define

\begin{equation}
A^\mu
:=
\eta^{\mu\nu}A_\nu.
\end{equation}

The field strength and the adjoint covariant derivative are defined by

\begin{equation}
F_{\mu\nu}
:=
i[A_\mu,A_\nu],
\end{equation}

and

\begin{equation}
D_\mu^X Y
:=
i[X_\mu,Y].
\end{equation}

Here \(\gamma^\mu_{\alpha\beta}\) denotes the ten-dimensional chiral gamma matrices acting on a Majorana--Weyl spinor. We use a fixed real basis of the positive-chirality Majorana--Weyl space

\begin{equation}
S_+\simeq\mathbb R^{16}.
\end{equation}

For the real symmetric chiral blocks used in the one-loop calculation, we write

\begin{equation}
\widehat\gamma^\mu:=i\gamma^\mu.
\end{equation}

The fermion operator can then be written as

\begin{equation}
\mathcal M_F
=
i\gamma^\mu D_\mu
=
\sum_\mu
\widehat\gamma^\mu\otimes D_\mu.
\end{equation}

The precise Clifford relations, matrix representation, and basis ordering used in the calculation are given in Appendix C.

The action is

\begin{equation}
S_{\mathrm{IIB}}[A,\psi]
=
\frac{1}{g^2}
\operatorname{Tr}
\left[
\frac14 F_{\mu\nu}F^{\mu\nu}
+
\frac{i}{2}
\psi^\alpha
\gamma^\mu_{\alpha\beta}
D_\mu^A\psi^\beta
\right].
\end{equation}

In this paper, we take

\begin{equation}
\psi^{(0)}:=0
\end{equation}

for the classical background and consider bosonic backgrounds \(A_\mu^{(0)}\). The corresponding bosonic action is

\begin{equation}
S_B[A]
:=
\frac{1}{4g^2}
\operatorname{Tr}
\left(
F_{\mu\nu}F^{\mu\nu}
\right).
\end{equation}

When it is useful to separate the coupling from the quartic quantity, we write

\begin{equation}
S_{B,0}[A]
:=
\frac14
\operatorname{Tr}
\left(
F_{\mu\nu}F^{\mu\nu}
\right),
\qquad
S_B[A]
=
\frac{S_{B,0}[A]}{g^2}.
\end{equation}

The classical phase is therefore

\begin{equation}
e^{iS_{B,0}[A]/g^2}
=
e^{iS_B[A]}.
\end{equation}

Unless otherwise stated, we use \(S_B\) as the manuscript quantity below.

For a fixed bosonic background \(A_\mu\), the fermionic action is quadratic in the Grassmann variables, so the fermionic integral is exact and gives the Pfaffian of the Majorana--Weyl fermion operator. We refer to this factor as the fermionic one-loop contribution in the background-field expansion; the full quantum theory still requires integration over the bosonic matrices, whose interactions generate higher-loop effects.

To characterize the spacetime extent of a matrix configuration, we introduce

\begin{equation}
G_{\mu\nu}[A]
:=
\frac{1}{N}
\operatorname{Tr}(A_\mu A_\nu).
\end{equation}

The components of \(G_{\mu\nu}\) depend on the choice of Lorentz frame.

Under a Lorentz transformation, \(G_{\mu\nu}\) transforms by congruence. Its positive, negative, and zero inertia is therefore frame independent by Sylvester's law of inertia. To obtain more detailed frame-independent quadratic information, we define the \textbf{Lorentzian extent operator}

\begin{equation}
K^\mu{}_\nu[A]
:=
\eta^{\mu\rho}G_{\rho\nu}[A].
\end{equation}

Mathematically, \(K^\mu{}_\nu\) is the mixed Lorentz tensor obtained by raising one index of \(G_{\mu\nu}\) with the Lorentz metric.

Under a Lorentz transformation, \(K^\mu{}_\nu\) transforms by similarity. Its eigenvalues are therefore independent of the Lorentz frame.

We denote the ten eigenvalues of \(K^\mu{}_\nu\) by

\begin{equation}
\lambda_1^K(A),\ldots,\lambda_{10}^K(A),
\end{equation}

and define

\begin{equation}
\mathcal S_K(A)
:=
\{
\lambda_1^K(A),\ldots,\lambda_{10}^K(A)
\}.
\end{equation}

We refer to this multiset, including multiplicities, as the \(K\)-spectrum.

Thus, if two matrix configurations \(A\) and \(A'\) satisfy

\begin{equation}
\mathcal S_K(A)
=
\mathcal S_K(A'),
\end{equation}

they have the same spectral data at the level of this coarse quadratic geometry.

The quantities \(G_{\mu\nu}\) and \(K^\mu{}_\nu\), however, are quadratic quantities obtained after tracing over the color-space information contained in the original matrices \(A_\mu\). The \(K\)-spectrum therefore need not determine the microscopic matrix configuration uniquely.

This leads to two questions. First, can physical configurations exist in which the microscopic noncommutative structure changes continuously while the coarse quadratic spectral data remain unchanged? Second, if such configurations exist, can one-loop quantum fluctuations detect differences that are invisible in the coarse quadratic geometry?

\subsection{\texorpdfstring{SO(5) vector-representation construction at \(N=5\)}{SO(5) vector-representation construction at N=5}}\label{so5-vector-representation-construction-at-n5}

The quantities \(G_{\mu\nu}\) and \(K^\mu{}_\nu\) compress the color-space information of the matrix configuration into quadratic trace data. To keep track of microscopic information that is lost in this compression, we use the representation structure of SO(5).

The IIB matrix model has ten bosonic matrices \(A_\mu\), while

\begin{equation}
\dim\mathfrak{so}(5)=10.
\end{equation}

This matching allows us to construct a finite-dimensional background ansatz in which the ten matrix directions are related to the generators of SO(5). This is one practical reason for using SO(5).

Let

\begin{equation}
V_1\simeq\mathbb R^5
\end{equation}

be the defining vector representation of SO(5). We set the matrix size of the IIB matrix model to \(N=5\) and identify the five-dimensional color space with \(V_1\).

Accordingly, we focus on the traceless matrix degrees of freedom. For \(N=5\),

\begin{equation}
\operatorname{End}(V_1)
\simeq
\mathbf 1
\oplus
\mathbf{10}
\oplus
\mathbf{14}.
\end{equation}

Here \(\operatorname{End}(V_1)\) is the full space of linear maps on \(V_1\), or equivalently the space of \(5\times5\) matrices. The three terms correspond to the scalar, antisymmetric, and symmetric traceless parts.

After removing the trace part,

\begin{equation}
\operatorname{End}_0(V_1)
\simeq
\mathbf{10}\oplus\mathbf{14}.
\end{equation}

The \(\mathbf{10}\) is the adjoint representation,

\begin{equation}
\mathbf{10}
\simeq
\Lambda^2V_1,
\end{equation}

while the \(\mathbf{14}\) is the symmetric traceless rank-two representation,

\begin{equation}
\mathbf{14}
\simeq
\operatorname{Sym}_0^2V_1.
\end{equation}

For Hermitian \(5\times5\) matrices, the \(\mathbf{10}\) sector is represented by \(i\) times real antisymmetric matrices, while the \(\mathbf{14}\) sector is represented by real symmetric traceless matrices.

The dimensions satisfy

\begin{equation}
10+14=24=5^2-1,
\end{equation}

so these two irreducible sectors exhaust the traceless \(5\times5\) matrix space.

We choose an SO(5) embedding into color space,

\begin{equation}
\iota:
\mathfrak{so}(5)\hookrightarrow\mathfrak{su}(5),
\end{equation}

and refer to the representation structure specified by this embedding as the \textbf{SO(5) structural frame}. Once this frame is fixed, each traceless Hermitian matrix \(A_\mu\) has a unique decomposition

\begin{equation}
A_\mu
=
A_\mu^{(10)}
+
A_\mu^{(14)},
\end{equation}

where \(A_\mu^{(10)}\) and \(A_\mu^{(14)}\) belong to the \(\mathbf{10}\) and \(\mathbf{14}\) sectors, respectively.

This decomposition allows us to track the two sectors separately and to study changes in the microscopic matrix structure that are not visible in the quadratic trace data. The explicit construction and the representation-theoretic details are given in Appendix A.

At \(N=5\), the one-loop fluctuation operators can also be constructed directly as finite-dimensional matrices. We can therefore study the SO(5)-resolved microscopic structure and the one-loop quantum response within the same finite-dimensional setup.

\subsection{Information retained and lost by the coarse geometry}\label{information-retained-and-lost-by-the-coarse-geometry}

Using the SO(5) decomposition introduced above, the orthogonality of the \(\mathbf{10}\) and \(\mathbf{14}\) sectors gives

\begin{equation}
G_{\mu\nu}[A]
=
\frac{1}{N}
\operatorname{Tr}
\left(
A_\mu^{(10)}A_\nu^{(10)}
\right)
+
\frac{1}{N}
\operatorname{Tr}
\left(
A_\mu^{(14)}A_\nu^{(14)}
\right).
\end{equation}

Thus, \(G_{\mu\nu}\), and therefore \(K^\mu{}_\nu\) and the \(K\)-spectrum, contain the combined quadratic contributions of the two SO(5) sectors. They do not keep the two contributions separately.

The contributions from the \(\mathbf{10}\) and \(\mathbf{14}\) sectors can therefore change while their sum remains unchanged. Such a change cannot be detected from \(G_{\mu\nu}\) alone. More generally, even when the \(K\)-spectrum is common, the SO(5)-resolved content and the noncommutative structure of the original matrices need not be the same.

This is why we distinguish between coarse quadratic geometry and microscopic matrix structure. In the next section, we show numerically that this situation occurs over the analyzed local interval of the constrained classical solution space: configurations that cannot be distinguished by the coarse quadratic spectral data are connected by a physical modulus.

\subsection{Constrained classical problem at fixed overall scale}\label{constrained-classical-problem-at-fixed-overall-scale}

When comparing different matrix configurations, we need to separate an overall rescaling from changes in shape and microscopic noncommutative structure. We therefore fix a Lorentzian quadratic quantity that scales quadratically under a homogeneous radial rescaling.

We define

\begin{equation}
Q_\rho[A]
:=
\operatorname{Tr}(A_\mu A^\mu)-Q_{\rm ref},
\qquad
Q_{\rm ref}:=6+2\sqrt3.
\end{equation}

Here \(Q_{\rm ref}\) is a fixed reference offset in the background parametrization and is not varied dynamically. The quantity

\begin{equation}
\operatorname{Tr}(A_\mu A^\mu)
=
\eta^{\mu\nu}\operatorname{Tr}(A_\mu A_\nu)
\end{equation}

is a Lorentzian quadratic contraction. It is not a positive-definite matrix norm.

We impose

\begin{equation}
Q_\rho[A]=9.
\end{equation}

This condition fixes the homogeneous radial normalization and defines a slice on which changes in shape and microscopic noncommutative structure can be compared. The value \(Q_\rho=9\) is used as a normalization slice for the analysis; we do not assume that this value is dynamically preferred. In Section 7 and Appendix F, we use a nearby radial continuation to check how the analysis extends away from this slice.

For the classical background, we take

\begin{equation}
\psi^{(0)}=0
\end{equation}

and look for stationary points of the bosonic action \(S_B\) on the fixed-\(Q_\rho\) hypersurface. Introducing a Lagrange multiplier \(\lambda_Q\), we obtain the constrained classical equation

\begin{equation}
[A^\nu,[A_\nu,A_\mu]]
=
2\lambda_Q A_\mu.
\end{equation}

The overall factor \(1/g^2\) in \(S_B\) can be absorbed into the definition of \(\lambda_Q\), so it does not appear explicitly in this equation.

Within this constrained classical problem, there is one continuous physical modulus that cannot be removed by gauge transformations or Lorentz-frame transformations. In the next section, we construct this modulus and study its physical properties.

\section{\texorpdfstring{\(N=5\) noncommutative modulus with common coarse geometry}{N=5 noncommutative modulus with common coarse geometry}}\label{n5-noncommutative-modulus-with-common-coarse-geometry}

\subsection{Constrained solutions with common coarse geometry}\label{constrained-solutions-with-common-coarse-geometry}

The fixed-\(Q_\rho\) constrained classical problem introduced in Section 2 admits a numerically continued local family of \(N=5\) solutions. For an exact constrained solution family at fixed \(Q_\rho\), constancy of the classical action follows analytically from stationarity. The accepted numerical backgrounds realize this identity to the precision quoted below, while the full \(K\)-spectrum remains common within numerical precision over the analyzed local interval.

We construct this family by numerically continuing solutions of the constrained classical equation

\begin{equation}
[A^\nu,[A_\nu,A_\mu]]
=
2\lambda_Q A_\mu
\end{equation}

subject to

\begin{equation}
Q_\rho[A]=9.
\end{equation}

Over the local interval analyzed below, both the constraint and the classical equation are satisfied at each point of the continuation.

The quantity \(Q_\rho\) is constant by construction. In addition, at a constrained stationary point,

\begin{equation}
dS_B
=
\lambda_Q\,dQ_\rho.
\end{equation}

Therefore, along a family with

\begin{equation}
dQ_\rho=0,
\end{equation}

we have

\begin{equation}
dS_B=0.
\end{equation}

Thus, for an exact constrained solution family on the fixed-\(Q_\rho\) slice, the bosonic classical action is constant. The numerical continuation below realizes this stationary-family identity to the quoted residual precision.

For this \(N=5\) family, the Lorentzian extent operator

\begin{equation}
K^\mu{}_\nu[A]
=
\eta^{\mu\rho}G_{\rho\nu}[A]
\end{equation}

also satisfies the exact scalar relation

\begin{equation}
Q_\rho
=
5K^\mu{}_\mu
-
(6+2\sqrt3).
\end{equation}

Thus, \(K^\mu{}_\mu\) is fixed exactly on the fixed-\(Q_\rho\) slice.

The constancy of the full \(K\)-spectrum is a numerical result rather than an analytic identity. Over the analyzed local interval, all ten eigenvalues

\begin{equation}
\lambda_i^K(A),
\qquad
i=1,\ldots,10,
\end{equation}

remain unchanged within numerical precision. Details of the numerical continuation, the residuals of the classical equation, and the variation of the \(K\)-spectrum are given in Appendix B; the numerical eigenvalue deviations are shown in Fig.~\ref{fig:k-spectrum-deviations}.

Thus, for an exact constrained solution family, \(Q_\rho\) and \(S_B\) are constant by construction and by the stationary identity. The accepted numerical backgrounds realize these equalities to the quoted precision, while the full \(K\)-spectrum is common within numerical precision over the analyzed local interval.

We next determine how many continuous physical directions remain after removing gauge and Lorentz-frame redundancies.

\subsection{One-dimensional physical modulus}\label{one-dimensional-physical-modulus}

The continuous solution family contains both physical deformations and redundant directions generated by gauge and Lorentz-frame transformations. We therefore remove these redundancies and examine the remaining physical fluctuation space.

For \(N=5\), one traceless Hermitian \(5\times5\) matrix has

\begin{equation}
5^2-1=24
\end{equation}

real degrees of freedom. The ten bosonic matrices therefore give a \(240\)-dimensional configuration space,

\begin{equation}
10(5^2-1)=240.
\end{equation}

Infinitesimal color-gauge transformations are generated by

\begin{equation}
\delta_\xi A_\mu
=
i[\xi,A_\mu],
\qquad
\xi\in\mathfrak{su}(5).
\end{equation}

Along the branch analyzed here, the gauge directions have rank \(24\). Removing them leaves

\begin{equation}
240-24=216
\end{equation}

directions.

We then impose the condition

\begin{equation}
Q_\rho[A]=9.
\end{equation}

This gives one independent local condition, so the tangent space to the fixed-\(Q_\rho\) gauge quotient has dimension

\begin{equation}
216-1=215.
\end{equation}

This space still contains changes of Lorentz frame. Infinitesimal Lorentz transformations act as

\begin{equation}
\delta_\omega A_\mu
=
\omega_\mu{}^\nu A_\nu,
\qquad
\omega_{\mu\nu}=-\omega_{\nu\mu}.
\end{equation}

On the branch considered here, the projected Lorentz-frame tangent map has rank \(45\). Removing these directions leaves a \(170\)-dimensional physical fluctuation space,

\begin{equation}
215-45=170.
\end{equation}

We now examine the quadratic variation of the constrained action on this space. We define the constrained Lagrange function by

\begin{equation}
\mathcal L_Q[A,\lambda_Q]
:=
S_B[A]
-
\lambda_Q\bigl(Q_\rho[A]-9\bigr),
\end{equation}

and denote its Hessian restricted to the \(170\)-dimensional physical fluctuation space by \(H_{\rm phys}\).

The inertia of this Hessian is

\begin{equation}
(95,74,1),
\end{equation}

corresponding to \(95\) positive eigenvalues, \(74\) negative eigenvalues, and one zero eigenvalue. Thus, \(169\) physical directions have nonzero quadratic curvature, while one direction remains as a zero mode.

A zero mode of the Hessian does not by itself imply a finite family of classical solutions. We therefore start from this direction and numerically continue the full constrained classical equation in both directions. This produces a continuous family of constrained stationary solutions over a local interval. The tangent to the resulting solution curve agrees with the zero mode of the physical Hessian.

The zero mode therefore extends to an actual one-dimensional family of classical solutions. We call this continuous physical degree of freedom the \textbf{physical modulus}.

We denote a local coordinate along the modulus by \(u\) and write the corresponding background as

\begin{equation}
A_\mu(u).
\end{equation}

We use the term \emph{accepted background} for a numerically continued solution that passes the predeclared floating-point checks on the constrained-equation residuals, the relevant ranks, and the nonzero spectral gaps. This terminology does not denote an interval-arithmetic or rigorous enclosure proof. The numerical thresholds, pointwise data, replay code, and validators are provided in the supplementary reproducibility archive.

The reference background is chosen at

\begin{equation}
u=0.
\end{equation}

The detailed dimension and rank counting, the physical Hessian, and the nonlinear continuation are given in Appendix B.

We next compare the backgrounds along this physical modulus using higher-order invariants.

\subsection{Variation of the microscopic noncommutative structure}\label{variation-of-the-microscopic-noncommutative-structure}

We now compare the classical matrix configurations

\begin{equation}
A_\mu=A_\mu(u)
\end{equation}

along the physical modulus before including quantum fluctuations.

Along this family,

\begin{equation}
Q_\rho[A(u)]=9,
\qquad
S_B[A(u)]=S_B[A(0)]
\end{equation}

hold for an exact constrained family. The accepted numerical backgrounds realize these equalities to the stated precision. The \(K\)-spectrum is also common within numerical precision over the analyzed local interval. At the level of the classical action and the coarse quadratic spectral data used in this paper, these backgrounds are therefore indistinguishable.

The matrix configuration, however, contains information beyond quadratic traces. For example,

\begin{equation}
G_{\mu\nu}
=
\frac1N\operatorname{Tr}(A_\mu A_\nu)
\end{equation}

contains quadratic matrix correlations but does not retain the full ordering information of products of noncommuting matrices.

To probe this information, we define

\begin{equation}
Q_2[A]
:=
\operatorname{Tr}(A_\mu A^\mu)
=
Q_\rho[A]+Q_{\rm ref}.
\end{equation}

Here \(Q_2\) is a Lorentzian quadratic contraction, not a positive-definite norm. We use it only as a homogeneous normalization factor and consider the sixth-order invariant

\begin{equation}
I_{6,\triangle}[A]
:=
\frac{1}{NQ_2[A]^3}
\operatorname{Tr}
\left(
A_\mu A_\nu A_\rho
A^\mu A^\nu A^\rho
\right).
\end{equation}

This quantity depends on the ordering of the matrices. Cyclic permutations of the full product are equivalent under the trace, but general permutations of noncommuting matrices are not. The invariant \(I_{6,\triangle}\) is therefore sensitive to information that is not contained in \(G_{\mu\nu}\) or \(K^\mu{}_\nu\).

We find that

\begin{equation}
I_{6,\triangle}[A(u)]
\end{equation}

varies nontrivially with \(u\), even though the \(K\)-spectrum remains common within numerical precision over the same local interval. The same conclusion is obtained from independent gauge- and Lorentz-invariant quantities constructed from the commutators

\begin{equation}
F_{\mu\nu}
=
i[A_\mu,A_\nu].
\end{equation}

The numerical variation of these higher-order invariants is given in Appendix B.

The backgrounds at different values of \(u\) are therefore not related simply by gauge transformations or changes of Lorentz frame. Their coarse quadratic geometry is the same within the accuracy described above, but their microscopic noncommutative structure is different.

This result can also be viewed through the SO(5) decomposition

\begin{equation}
A_\mu
=
A_\mu^{(10)}
+
A_\mu^{(14)}.
\end{equation}

The tensor \(G_{\mu\nu}\) contains the combined quadratic contributions from the \(\mathbf{10}\) and \(\mathbf{14}\) sectors, but it does not retain all the matrix-ordering and commutator information contained in each sector.

The physical modulus therefore gives a continuous classical family with

\begin{equation}
\begin{array}{c}
\text{the same classical action and a common }K\text{-spectrum}\\
\text{within numerical precision over the analyzed local interval},\\[1mm]
\text{but different microscopic noncommutative structures}.
\end{array}
\end{equation}

We have thus numerically continued a local family of classical backgrounds that cannot be distinguished by the coarse quadratic spectral data used here within the stated precision but differ at the microscopic matrix level. In the next section, we ask whether one-loop quantum fluctuations can distinguish these backgrounds.

\section{One-loop quantum resolution of the classical degeneracy}\label{one-loop-quantum-resolution-of-the-classical-degeneracy}

\subsection{One-loop expansion around the physical modulus}\label{one-loop-expansion-around-the-physical-modulus}

In Section 3, we found a one-dimensional physical modulus in the fixed-\(Q_\rho\) constrained classical problem. Along this modulus, the classical action is constant and the full \(K\)-spectrum remains common within numerical precision over the analyzed local interval, while the microscopic noncommutative structure changes.

We write the classical backgrounds as

\begin{equation}
A_\mu=A_\mu(u),
\end{equation}

where \(u\) is the local coordinate along the physical modulus. In this section, we include quantum fluctuations around each background and ask whether they distinguish configurations that have the same classical action and the same coarse quadratic spectral data.

We expand the bosonic matrices as

\begin{equation}
A_\mu=A_\mu(u)+a_\mu,
\end{equation}

where \(a_\mu\) denotes the bosonic fluctuation around the background \(A_\mu(u)\). We also include Majorana--Weyl fermionic fluctuations around the same background.

At one loop, we expand the action to quadratic order in the fluctuations and perform the corresponding Gaussian integrations. Since \(N=5\), all quadratic fluctuation operators are finite-dimensional matrices.

Not all components of \(a_\mu\) represent independent physical fluctuations. Color-gauge redundancy is treated by background gauge fixing together with the corresponding Faddeev--Popov ghosts. The direction normal to the fixed-\(Q_\rho\) surface and the directions corresponding to changes of Lorentz frame are also removed from the physical bosonic fluctuation space.

As shown in Section 3, the physical bosonic fluctuation space has dimension \(170\). Its constrained Hessian contains one zero mode, so that

\begin{equation}
170
=
1_{\rm modulus}
+
169_{\rm transverse}.
\end{equation}

The zero mode is tangent to the classical solution family \(A_\mu(u)\). We keep this direction as the collective coordinate \(u\).

The remaining \(169\) directions have nonzero quadratic curvature. We refer to this physical transverse subspace as the \textbf{hard sector}. Here ``hard'' only means that these directions have a nonzero quadratic response, in contrast to the modulus zero mode.

The one-loop calculation naturally separates three kinds of quantities. First, the Majorana--Weyl sector gives the Pfaffian magnitude, from which we define an intrinsic scalar response on the gauge/proper-Lorentz quotient. Second, after choosing a Lorentz section and the auxiliary coefficient-space metric used to define the hard basis, the bosonic, ghost, and fermionic determinants can be combined into a useful fixed-section diagnostic,

\begin{equation}
\Gamma_{1,\mathrm{sec}}^{\det}(u).
\end{equation}

Third, the physical modulus supplies the collective-coordinate measure factor

\begin{equation}
J_{\rm mod}(u),
\end{equation}

and the complete quotient construction gives the relative local one-loop density coefficient

\begin{equation}
W_{\rm rel}(u)
\end{equation}

in the fixed continuation coordinate \(u\). The geometric object associated with this last quantity is the density one-form. The fixed-section diagnostic is not used as an intrinsic scalar on the Lorentz quotient.

\subsection{Quadratic fluctuation operators and Gaussian factors}\label{quadratic-fluctuation-operators-and-gaussian-factors}

Around each classical background \(A_\mu(u)\), we restrict the second variation of the constrained Lagrange function

\begin{equation}
\mathcal L_Q[A,\lambda_Q]
=
S_B[A]
-
\lambda_Q\bigl(Q_\rho[A]-9\bigr)
\end{equation}

to the \(170\)-dimensional physical fluctuation space. We denote the resulting Hessian by

\begin{equation}
H_{\rm phys}(u).
\end{equation}

To define finite-dimensional bases, projections, and local Jacobians, we introduce an auxiliary positive-definite inner product on the real coefficient space of the bosonic matrix components,

\begin{equation}
\langle a,b\rangle_{\rm coeff}
:=
\delta^{\mu\nu}
\operatorname{Tr}(a_\mu b_\nu).
\end{equation}

This inner product is used only for normalization and orthogonal decomposition in coefficient space. It is different from the Lorentzian metric \(\eta_{\mu\nu}\) that enters the action and the constraint.

Let \(t(u)\) be the normalized tangent to the physical modulus, and let \(U_h(u)\) be an orthonormal basis of the \(169\)-dimensional subspace orthogonal to \(t(u)\),

\begin{equation}
U_h(u)^{\mathsf T}U_h(u)=I_{169},
\qquad
U_h(u)^{\mathsf T}t(u)=0.
\end{equation}

The hard-sector Hessian is then

\begin{equation}
H_{\rm hard}(u)
=
U_h(u)^{\mathsf T}
H_{\rm phys}(u)
U_h(u).
\end{equation}

This operator describes the nonzero quadratic response of the \(169\) physical transverse bosonic fluctuations.

The Lorentzian bosonic Gaussian is defined locally around each real background by adding an infinitesimal damping term in the real fluctuation coefficient space and taking the limit as the damping is removed. The details are given in Appendix D. Along the analyzed branch, the inertia of \(H_{\rm hard}(u)\) is constant,

\begin{equation}
(95,74).
\end{equation}

The corresponding bosonic Gaussian phase is therefore also common along the branch,

\begin{equation}
e^{i\phi_B}
=
\exp\left[
\frac{i\pi}{4}(95-74)
\right]
=
e^{-3\pi i/4}.
\end{equation}

Color-gauge redundancy is treated in background gauge. With

\begin{equation}
D_\mu X
:=
i[A_\mu(u),X],
\end{equation}

we impose

\begin{equation}
D^\mu a_\mu=0
\end{equation}

and define the Faddeev--Popov operator by

\begin{equation}
\Delta_{\rm FP}(u)
=
-D_\mu D^\mu.
\end{equation}

For \(N=5\), this operator acts on the \(24\)-dimensional interacting adjoint color space. The ghost Gaussian integral gives

\begin{equation}
\det\Delta_{\rm FP}(u).
\end{equation}

For the fermionic sector, we use the Majorana--Weyl quadratic operator

\begin{equation}
\mathcal M_F(u)
=
i\gamma^\mu D_\mu.
\end{equation}

It acts on the product of the real \(16\)-dimensional Majorana--Weyl spinor space and the \(24\)-dimensional adjoint color space, giving a \(384\)-dimensional real fluctuation space. In the fixed real basis used throughout the calculation, \(\mathcal M_F(u)\) is real and antisymmetric, and the Grassmann Gaussian integral gives

\begin{equation}
\operatorname{Pf}\mathcal M_F(u).
\end{equation}

The precise gauge-fixing convention, gamma-matrix representation, basis ordering, and Pfaffian orientation are given in Appendix C.

Up to an overall normalization independent of \(u\), the Gaussian one-loop factor before including the collective-coordinate measure is

\begin{equation}
Z_{\rm det}^{(1)}(u)
\propto
e^{i\phi_B}
\frac{
\det\Delta_{\rm FP}(u)\,
\operatorname{Pf}\mathcal M_F(u)
}{
\sqrt{|\det H_{\rm hard}(u)|}
}.
\end{equation}

In the chosen Lorentz section and auxiliary coefficient-space normalization, we define the combined determinant/Pfaffian diagnostic

\begin{equation}
\Gamma_{1,\mathrm{sec}}^{\det}(u)
=
\frac12\log|\det H_{\rm hard}(u)|
-
\log|\det\Delta_{\rm FP}(u)|
-
\log|\operatorname{Pf}\mathcal M_F(u)|.
\end{equation}

This quantity contains the magnitude contribution from the transverse bosonic, ghost, and fermionic quadratic fluctuations. It does not include the common Lorentzian bosonic phase or the collective-coordinate measure. Because the orthonormal hard basis is defined with the auxiliary positive-definite coefficient-space metric, \(\Gamma_{1,\mathrm{sec}}^{\det}\) is a fixed-section diagnostic: it is independent of the parametrization of the branch in that section, but it is not a scalar under a change of proper-Lorentz representative.

Relative to the reference background \(u=0\), we define

\begin{equation}
\Delta\Gamma_{1,\mathrm{sec}}^{\det}(u)
=
\Gamma_{1,\mathrm{sec}}^{\det}(u)
-
\Gamma_{1,\mathrm{sec}}^{\det}(0).
\end{equation}

The intrinsic scalar quantity used below is instead the Majorana--Weyl Pfaffian magnitude.

\subsection{Intrinsic Pfaffian response and fixed-section determinant diagnostic}\label{one-loop-determinant-contribution-along-the-modulus}

For the fermionic sector, define

\begin{equation}
\mathcal P_F(u)
:=
\log|\operatorname{Pf}\mathcal M_F(u)|,
\end{equation}

and the root-referenced response

\begin{equation}
\Delta\mathcal P_F(u)
:=
\mathcal P_F(u)-\mathcal P_F(0).
\end{equation}

Appendix C shows that \(|\operatorname{Pf}\mathcal M_F|\) is invariant under color-gauge transformations and connected proper-Lorentz transformations on the interacting adjoint/Majorana--Weyl fluctuation space. Thus \(\Delta\mathcal P_F\) is a scalar function on the local physical quotient branch.

On the \(23\) accepted backgrounds, \(\Delta\mathcal P_F(u)\) is nonconstant. Its sampled logarithmic span is

\begin{equation}
\operatorname{span}_u\Delta\mathcal P_F
\simeq
0.05072,
\end{equation}

so that the relative Pfaffian magnitude

\begin{equation}
\mathcal R_F(u)
:=
\exp[\Delta\mathcal P_F(u)]
\end{equation}

has sampled maximum-to-minimum ratio

\begin{equation}
\frac{\max\mathcal R_F}{\min\mathcal R_F}
\simeq
1.0520.
\end{equation}

This already distinguishes points on the physical modulus at one-loop order, even though the classical action is common and the full \(K\)-spectrum remains common within numerical precision over the analyzed local interval.

For completeness, the fixed-section combined determinant diagnostic also varies. On the same backgrounds,

\begin{equation}
\operatorname{span}_u
\bigl[-\Delta\Gamma_{1,\mathrm{sec}}^{\det}\bigr]
\simeq
0.276,
\end{equation}

corresponding to a sampled maximum-to-minimum factor of approximately \(1.32\). The largest contribution to this fixed-section variation comes from the hard bosonic determinant; the Pfaffian gives a smaller but intrinsic scalar variation, while the Faddeev--Popov determinant is numerically constant to the quoted precision. We report the fixed-section decomposition because it is useful for understanding the finite-dimensional Gaussian calculation, but we do not interpret it as an intrinsic scalar on the Lorentz quotient.

The complete boson--ghost--fermion result must be combined with the quotient measure. We turn to that density one-form next.

\subsection{Collective-coordinate measure and local one-loop density}\label{collective-coordinate-measure-and-local-one-loop-density}

The zero mode of the physical Hessian is separated from the Gaussian integration and is kept as the collective coordinate \(u\) along the classical solution curve

\begin{equation}
u\longmapsto A_\mu(u).
\end{equation}

Using the auxiliary coefficient-space inner product introduced in Section 4.2, we define

\begin{equation}
G_{uu}(u)
=
\left\langle
\frac{dA}{du},
\frac{dA}{du}
\right\rangle_{\rm coeff},
\end{equation}

so that the line element along the solution curve is

\begin{equation}
d\ell
=
\sqrt{G_{uu}(u)}\,du.
\end{equation}

The quantity \(G_{uu}\) is an auxiliary metric on the finite-dimensional coefficient space. It is not the Lorentzian spacetime metric.

The measure on the physical solution curve also contains local Jacobian factors associated with the restriction to the fixed-\(Q_\rho\) surface and with the removal of color-gauge and Lorentz-frame orbits. We denote these factors by

\begin{equation}
J_Q(u),
\qquad
J_G(u),
\qquad
J_L(u).
\end{equation}

We use the convention in which the Faddeev--Popov determinant remains explicitly in the Gaussian factor. The collective-coordinate factor consistent with this convention is

\begin{equation}
J_{\rm mod}(u)
=
\sqrt{G_{uu}(u)}
\frac{
J_G(u)J_L(u)
}{
J_Q(u)\det\Delta_{\rm FP}(u)
}.
\end{equation}

With this definition,

\begin{equation}
J_{\rm mod}(u)\det\Delta_{\rm FP}(u)
=
\sqrt{G_{uu}(u)}
\frac{
J_G(u)J_L(u)
}{
J_Q(u)
},
\end{equation}

so that the gauge-orbit contribution is counted only once. The construction of these Jacobians is given in Appendix D.

On the fixed-\(Q_\rho=9\) slice, we define the local one-loop density magnitude by

\begin{equation}
\mathcal W_{\rm 1loop}(u,9)
=
J_{\rm mod}(u)e^{-\Gamma_{1,\mathrm{sec}}^{\det}(u)}.
\end{equation}

Equivalently,

\begin{equation}
\mathcal W_{\rm 1loop}(u,9)
=
J_{\rm mod}(u)
\frac{
|\det\Delta_{\rm FP}(u)|\,
|\operatorname{Pf}\mathcal M_F(u)|
}{
\sqrt{|\det H_{\rm hard}(u)|}
}.
\end{equation}

Restoring the classical phase and the Lorentzian bosonic Gaussian phase, the local contribution along the physical modulus takes the form

\begin{equation}
\mathcal N\,
e^{iS_B[A(u)]}
e^{i\phi_B}
\mathcal W_{\rm 1loop}(u,9)\,du.
\end{equation}

Both \(S_B[A(u)]\) and \(\phi_B\) are common along the analyzed modulus. They therefore cancel in relative magnitude comparisons.

We define the relative local one-loop density magnitude with respect to the reference background \(u=0\) by

\begin{equation}
W_{\rm rel}(u)
:=
\frac{
\mathcal W_{\rm 1loop}(u,9)
}{
\mathcal W_{\rm 1loop}(0,9)
}.
\end{equation}

Thus,

\begin{equation}
W_{\rm rel}(u)
=
\frac{J_{\rm mod}(u)}{J_{\rm mod}(0)}
\exp[-\Delta\Gamma_{1,\mathrm{sec}}^{\det}(u)].
\end{equation}

The quantity that descends to the gauge/proper-Lorentz quotient is the complete density one-form. Under a change of Lorentz representative, the auxiliary coefficient-space hard determinant and geometric Jacobian need not be separately invariant, but their changes cancel in the full quotient density. Appendix D and the reproducibility supplement verify this cancellation explicitly for finite boosts. Under a reparametrization of the modulus, the coefficient transforms as a one-dimensional density. The numerical quantity \(W_{\rm rel}(u)\) is therefore only the coefficient ratio in the fixed continuation coordinate \(u\) defined in Section 3; its nonconstancy by itself is not an intrinsic scalar statement. The intrinsic scalar discrimination used in this paper is the Pfaffian response \(\Delta\mathcal P_F\) of Section 4.3.

As a useful geometric parametrization of the same density one-form, we also use the auxiliary coefficient-space proper length

\begin{equation}
d\ell
=
\sqrt{G_{uu}(u)}\,du.
\end{equation}

Once this auxiliary metric is fixed, the corresponding relative proper-length density coefficient is

\begin{equation}
W_{\rm rel}^{(\ell)}(u)
:=
W_{\rm rel}(u)
\sqrt{\frac{G_{uu}(0)}{G_{uu}(u)}}.
\end{equation}

This proper length is an auxiliary coefficient-space quantity, not a Lorentzian spacetime distance. On the same \(23\) sampled backgrounds, we find

\begin{equation}
0.8068
\lesssim
W_{\rm rel}^{(\ell)}(u)
\lesssim
1.0320,
\end{equation}

with sampled maximum-to-minimum ratio

\begin{equation}
\frac{\max W_{\rm rel}^{(\ell)}}{\min W_{\rm rel}^{(\ell)}}
\simeq
1.2790.
\end{equation}

\begin{figure}[t]
\centering
\includegraphics[width=0.86\textwidth]{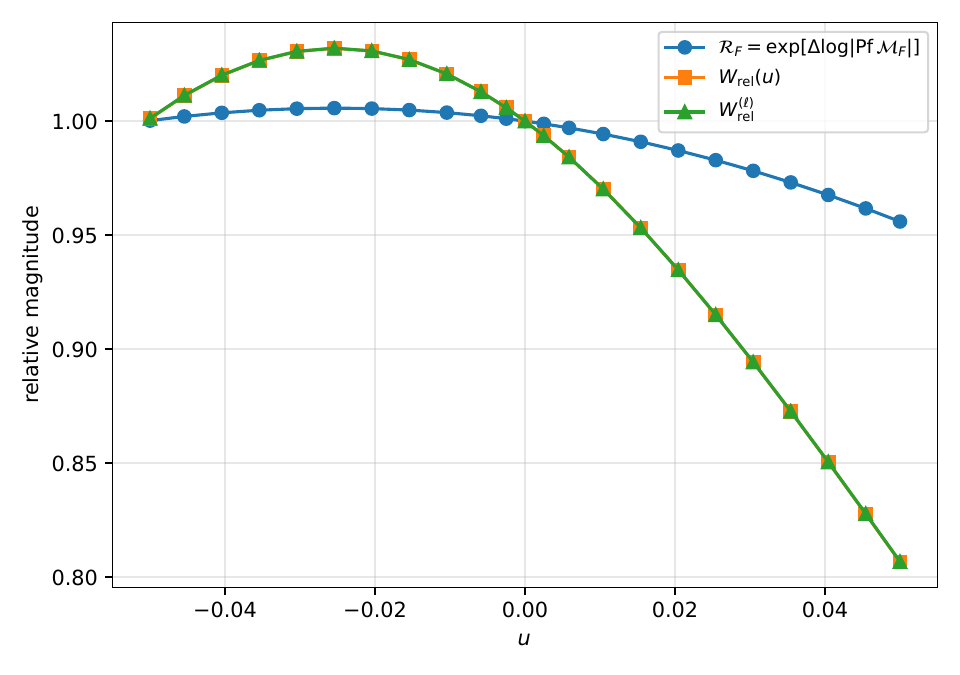}
\caption{One-loop quantities on the $23$ sampled backgrounds. The relative Pfaffian magnitude $\mathcal R_F(u)=\exp[\Delta\mathcal P_F(u)]$ is an intrinsic scalar under color-gauge and connected proper-Lorentz transformations. The coefficient $W_{\rm rel}(u)$ refers to the fixed continuation coordinate $u$, while $W_{\rm rel}^{(\ell)}$ is the coefficient of the same quotient density one-form with respect to the auxiliary coefficient-space proper length. Neither density coefficient is a normalized probability or an effective potential.}
\label{fig:wrel-profile}
\end{figure}

For comparison, the intrinsic Pfaffian response factor \(\mathcal R_F(u)\), the fixed-\(u\) density coefficient \(W_{\rm rel}(u)\), and the auxiliary proper-length coefficient \(W_{\rm rel}^{(\ell)}(u)\) are shown together in Fig.~\ref{fig:wrel-profile}. The Pfaffian factor has sampled maximum-to-minimum ratio approximately \(1.052\), while the two density coefficients have ratios approximately \(1.279\).

The microscopic noncommutative structure that is invisible to the coarse quadratic spectral data is therefore distinguished intrinsically by the fermionic Pfaffian response. The complete one-loop calculation supplies, in addition, a well-defined quotient density one-form; its numerical coefficient requires a chosen branch coordinate. The fixed-section hard determinant and its spectrum remain useful diagnostics of the local Gaussian calculation, but they are not treated as Lorentz-quotient scalars. The detailed Lorentzian Gaussian prescription, quotient Jacobians, and representative-covariance check are given in Appendix D.

\section{SO(5) representation probe of the microscopic structure}\label{so5-representation-probe-of-the-microscopic-structure}

\subsection{Canonical SO(5) representation family}\label{canonical-so5-representation-family}

In Sections 3 and 4, we studied a one-dimensional physical modulus

\begin{equation}
A_\mu(u)
\end{equation}

with a common classical action and a \(K\)-spectrum that remains common within numerical precision over the analyzed local interval. We found that the microscopic noncommutative structure changes along this modulus and that the same backgrounds are distinguished intrinsically by the Majorana--Weyl Pfaffian response at one-loop order. The complete local one-loop quotient measure gives a density one-form whose coefficient requires a specified branch parametrization.

For each \(N=5\) background on the physical modulus, we keep its microscopic data unchanged and examine them through different SO(5) representations. The purpose is to see which parts of the microscopic structure are visible in the ordinary \(N=5\) quadratic trace and which parts are hidden.

As discussed in Section 2, for the SO(5) vector representation

\begin{equation}
V_1\simeq\mathbb R^5,
\end{equation}

the traceless matrix space is decomposed as

\begin{equation}
\operatorname{End}_0(V_1)
=
\mathbf{10}\oplus\mathbf{14}.
\end{equation}

Once the SO(5) structural frame is fixed, each background on the physical modulus can be written as

\begin{equation}
A_\mu(u)
=
A_\mu^{(10)}(u)
+
A_\mu^{(14)}(u).
\end{equation}

Let

\begin{equation}
T_{ab}^{(1)},
\qquad
1\le a<b\le5,
\end{equation}

be the Hermitian SO(5) generators in the \(\mathbf{10}\) sector. We represent the \(\mathbf{14}\) sector by a symmetric traceless matrix

\begin{equation}
S_\mu(u)
\in
\operatorname{Sym}_0^2(\mathbb R^5).
\end{equation}

The \(N=5\) background can then be written as

\begin{equation}
A_\mu(u)
=
\sum_{a<b}
g_\mu^{ab}(u)T_{ab}^{(1)}
+
\mathcal E_1(S_\mu(u)),
\qquad
\mathcal E_1(S)=S.
\end{equation}

The coefficients \(g_\mu^{ab}(u)\) and the matrices \(S_\mu(u)\) contain the microscopic data in the \(\mathbf{10}\) and \(\mathbf{14}\) sectors, respectively.

For each integer \(L\ge1\), we use the standard symmetric traceless rank-\(L\) SO(5) representation \cite{Ramgoolam:2001,Kimura:2002,Sperling:2017},

\begin{equation}
V_L
=
\operatorname{Sym}_0^L(\mathbb R^5).
\end{equation}

Its dimension and quadratic Casimir are

\begin{equation}
N_L
=
\frac{(L+1)(L+2)(2L+3)}{6},
\end{equation}

and

\begin{equation}
C_2(L)=L(L+3).
\end{equation}

In particular,

\begin{equation}
N_1=5,
\qquad
N_2=14,
\qquad
N_3=30.
\end{equation}

Here \(N_L\) denotes the dimension of the representation \(V_L\), while the dynamical matrix size remains \(N=5\) throughout this paper.

For the \(\mathbf{10}\) sector, we use the same coefficients \(g_\mu^{ab}(u)\) and couple them to the SO(5) generators \(T_{ab}^{(L)}\) acting on \(V_L\).

For the \(\mathbf{14}\) sector, we define

\begin{equation}
\mathcal E_L(S)
=
\frac16
\sum_{a,b,c=1}^{5}
S_{ab}
\left\{
T_{ac}^{(L)},
T_{bc}^{(L)}
\right\}.
\end{equation}

The normalization is chosen so that

\begin{equation}
\mathcal E_1(S)=S.
\end{equation}

The SO(5) equivariance and the normalization of this map are discussed in Appendix A.

Using the same microscopic coefficient data, we then construct

\begin{equation}
\widetilde A_\mu^{(L)}(u)
=
\sum_{a<b}
g_\mu^{ab}(u)T_{ab}^{(L)}
+
\mathcal E_L\!\left(S_\mu(u)\right).
\end{equation}

Thus, the physical modulus and the underlying \(N=5\) microscopic data are unchanged. Only the SO(5) representation used to read these data is changed.

This construction is defined relative to the chosen SO(5) structural frame. Under a color-gauge transformation, the background and the structural frame are transported together. The resulting spectral quantities are therefore independent of the chosen gauge representative. The detailed covariance properties are given in Appendix E.

\subsection{Representation weights of the two SO(5) sectors}\label{representation-weights-of-the-two-so5-sectors}

We now separate the quadratic contributions of the \(\mathbf{10}\) and \(\mathbf{14}\) sectors.

For the \(\mathbf{10}\) sector, we define

\begin{equation}
\mathsf X_{\mu\nu}(u)
=
\sum_{a<b}
g_\mu^{ab}(u)g_\nu^{ab}(u).
\end{equation}

For the \(\mathbf{14}\) sector, we define

\begin{equation}
\mathsf Y_{\mu\nu}(u)
=
\operatorname{tr}_5
\left[
S_\mu(u)S_\nu(u)
\right].
\end{equation}

These tensors keep track of the quadratic contributions from the two sectors separately.

On \(V_L\), the trace normalization of the SO(5) generators is written as

\begin{equation}
\operatorname{Tr}_L
\left(
T_{ab}^{(L)}T_{cd}^{(L)}
\right)
=
\kappa_L
\left(
\delta_{ac}\delta_{bd}
-
\delta_{ad}\delta_{bc}
\right),
\end{equation}

with

\begin{equation}
\kappa_L
=
\frac{N_LC_2(L)}{10}.
\end{equation}

For the \(\mathbf{14}\) sector,

\begin{equation}
\operatorname{Tr}_L
\left[
\mathcal E_L(S)\mathcal E_L(S')
\right]
=
\beta_L
\operatorname{tr}_5(SS'),
\end{equation}

where

\begin{equation}
\beta_L
=
\frac{
N_LC_2(L)[4C_2(L)+5]
}{420}.
\end{equation}

The cross term between the \(\mathbf{10}\) and \(\mathbf{14}\) sectors vanishes under the trace. These relations are derived in Appendix A.

The relative quadratic-trace weight of the \(\mathbf{14}\) sector with respect to the \(\mathbf{10}\) sector is therefore

\begin{equation}
\rho_L
:=
\frac{\beta_L}{\kappa_L}
=
\frac{4L(L+3)+5}{42}.
\end{equation}

For the first three representations,

\begin{equation}
\rho_1=\frac12,
\qquad
\rho_2=\frac{15}{14},
\qquad
\rho_3=\frac{11}{6}.
\end{equation}

Moreover,

\begin{equation}
\rho_{L+1}-\rho_L
=
\frac{4(L+2)}{21}>0.
\end{equation}

Thus, the relative weight of the \(\mathbf{14}\) sector increases exactly with \(L\).

Using these trace relations, the unnormalized quadratic trace tensor becomes

\begin{equation}
\frac1{N_L}
\operatorname{Tr}_L
\left[
\widetilde A_\mu^{(L)}(u)
\widetilde A_\nu^{(L)}(u)
\right]
=
\frac{C_2(L)}{10}
\left[
\mathsf X_{\mu\nu}(u)
+
\rho_L\mathsf Y_{\mu\nu}(u)
\right].
\end{equation}

There are two kinds of \(L\)-dependence in this expression. The factor \(C_2(L)\) changes the overall scale, while \(\rho_L\) changes the relative weight of the two SO(5) sectors.

We are interested in the second effect. We therefore remove the overall representation-dependent scale before comparing different values of \(L\).

For the original \(N=5\) backgrounds, the fixed-\(Q_\rho\) condition gives

\begin{equation}
\operatorname{Tr}(A_\mu A^\mu)
=
9+Q_{\rm ref}
=
15+2\sqrt3.
\end{equation}

We write

\begin{equation}
D_0:=15+2\sqrt3.
\end{equation}

Using the same Lorentzian quadratic normalization for the higher-\(L\) construction, we define the normalized extent operator by

\begin{equation}
\bigl(K^{(L)}(u)\bigr)^\mu{}_{\nu}
=
\frac{D_0}
{
5\,\eta^{\rho\sigma}
[
\mathsf X_{\rho\sigma}(u)
+
\rho_L\mathsf Y_{\rho\sigma}(u)
]
}
\,
\eta^{\mu\lambda}
\left[
\mathsf X_{\lambda\nu}(u)
+
\rho_L\mathsf Y_{\lambda\nu}(u)
\right].
\end{equation}

The overall Casimir factor \(C_2(L)\) cancels in this normalization. The representation dependence of the spectral shape is therefore controlled by the relative weight \(\rho_L\).

\subsection{\texorpdfstring{Microscopic deformation hidden at \(L=1\)}{Microscopic deformation hidden at L=1}}\label{microscopic-deformation-hidden-at-l1}

We can now see why the physical modulus found in Section 3 is not visible in the ordinary \(N=5\) \(K\)-spectrum.

The case \(L=1\) corresponds to the actual \(N=5\) color space used in Sections 2--4, and

\begin{equation}
\rho_1=\frac12.
\end{equation}

The ordinary \(N=5\) quadratic extent tensor therefore reads the two SO(5) sectors through the combination

\begin{equation}
\mathsf X_{\mu\nu}
+
\frac12\mathsf Y_{\mu\nu}.
\end{equation}

Indeed,

\begin{equation}
G_{\mu\nu}
=
\frac25
\left[
\mathsf X_{\mu\nu}
+
\frac12\mathsf Y_{\mu\nu}
\right].
\end{equation}

Along the physical modulus, the quadratic contributions from the \(\mathbf{10}\) and \(\mathbf{14}\) sectors change separately. However, when they are combined with the \(L=1\) relative weight \(1:\frac12\), the resulting \(K\)-spectrum remains common within numerical precision over the analyzed local interval.

The microscopic redistribution between the two sectors is therefore already present at the classical level, but it is hidden by the particular combination used in the ordinary \(N=5\) quadratic trace.

For \(L>1\),

\begin{equation}
\rho_L\neq\frac12.
\end{equation}

The same microscopic data are then read through a different combination,

\begin{equation}
\mathsf X+\rho_L\mathsf Y.
\end{equation}

The sector redistribution that is hidden at \(L=1\) therefore becomes visible as a spectral response at higher \(L\).

To measure the change of the full spectrum, we define

\begin{equation}
I_L(u)
:=
\operatorname{Tr}_{10}
\left[
\bigl(K^{(L)}(u)\bigr)^2
\right].
\end{equation}

This quantity is the sum of the squares of the ten eigenvalues of \(K^{(L)}(u)\).

For the \(23\) accepted backgrounds, we define the sampled variation by

\begin{equation}
\operatorname{span}_u I_L
:=
\max_u I_L(u)-\min_u I_L(u).
\end{equation}

For \(L=1\), the variation of \(I_1(u)\) remains at the level of numerical precision. In contrast,

\begin{equation}
\operatorname{span}_u I_2
\simeq
1.83\times10^{-4},
\end{equation}

and

\begin{equation}
\operatorname{span}_u I_3
\simeq
5.57\times10^{-4}.
\end{equation}

Direct comparison of the eigenvalues of \(K^{(L)}(u)\) gives the same result. The \(L=1\) spectrum remains common within numerical precision, while clear spectral variations appear for \(L=2\) and \(L=3\).

The higher-\(L\) construction reads the same classical microscopic deformation with a different relative weight of the two sectors.

\subsection{\texorpdfstring{The \(3+6\) ordered spectral hierarchy across \(L\)}{The 3+6 ordered spectral hierarchy across L}}\label{the-36-ordered-spectral-hierarchy-across-l}

The higher-\(L\) representation probe also reveals another feature of the spectrum.

At the reference background,

\begin{equation}
\bigl(K^{(L)}(0)\bigr)^\mu{}_{\nu}
\end{equation}

has one eigenvalue consistent with zero within numerical precision and nine positive eigenvalues. We order the positive eigenvalues as

\begin{equation}
0<
\lambda_1^{(L)}
\le
\lambda_2^{(L)}
\le
\cdots
\le
\lambda_9^{(L)}.
\end{equation}

For this reference background, the numerically identified zero-eigenvalue direction is timelike. Details are given in Appendix A.

The nine positive eigenvalues are not distributed uniformly. The largest three are larger, on average, than the remaining six. We measure this separation by

\begin{equation}
R_{3+6}(L,u)
=
\frac{
(\lambda_7^{(L)}(u)+\lambda_8^{(L)}(u)+\lambda_9^{(L)}(u))/3
}{
(\lambda_1^{(L)}(u)+\cdots+\lambda_6^{(L)}(u))/6
}.
\end{equation}

At the reference background, we use the shorthand

\begin{equation}
R_{3+6}(L):=R_{3+6}(L,0).
\end{equation}

The notation \(3+6\) refers to this ordered separation between the largest three and the remaining six positive eigenvalues. It does not imply an exact degeneracy.

For the reference background,

\begin{equation}
R_{3+6}(1)\simeq1.429.
\end{equation}

Thus, already at \(L=1\), the largest three positive eigenvalues are larger on average than the remaining six.

At higher representation levels, we find

\begin{equation}
R_{3+6}(2)\simeq1.574,
\end{equation}

and

\begin{equation}
R_{3+6}(3)\simeq1.763.
\end{equation}

Since

\begin{equation}
\rho_L\longrightarrow\infty
\qquad
(L\to\infty),
\end{equation}

the \(\mathbf{14}\) sector dominates the normalized spectral shape in the canonical large-\(L\) limit. In this limit,

\begin{equation}
\bigl(K^{(L)}\bigr)^\mu{}_{\nu}
\longrightarrow
(K_\infty)^\mu{}_{\nu}
=
\frac{D_0}
{
5\,\eta^{\rho\sigma}\mathsf Y_{\rho\sigma}
}
\,
\eta^{\mu\lambda}\mathsf Y_{\lambda\nu}.
\end{equation}

For the reference background,

\begin{equation}
R_{3+6}(\infty)\simeq5.946.
\end{equation}

The numerical sequence is therefore

\begin{equation}
1.429
\;\longrightarrow\;
1.574
\;\longrightarrow\;
1.763
\;\longrightarrow\;
5.946.
\end{equation}

The monotonic increase of \(\rho_L\) is an exact representation-theoretic result. The values of \(R_{3+6}(L)\), however, are numerical results for the reference background.

\begin{figure}[t]
\centering
\includegraphics[width=0.82\textwidth]{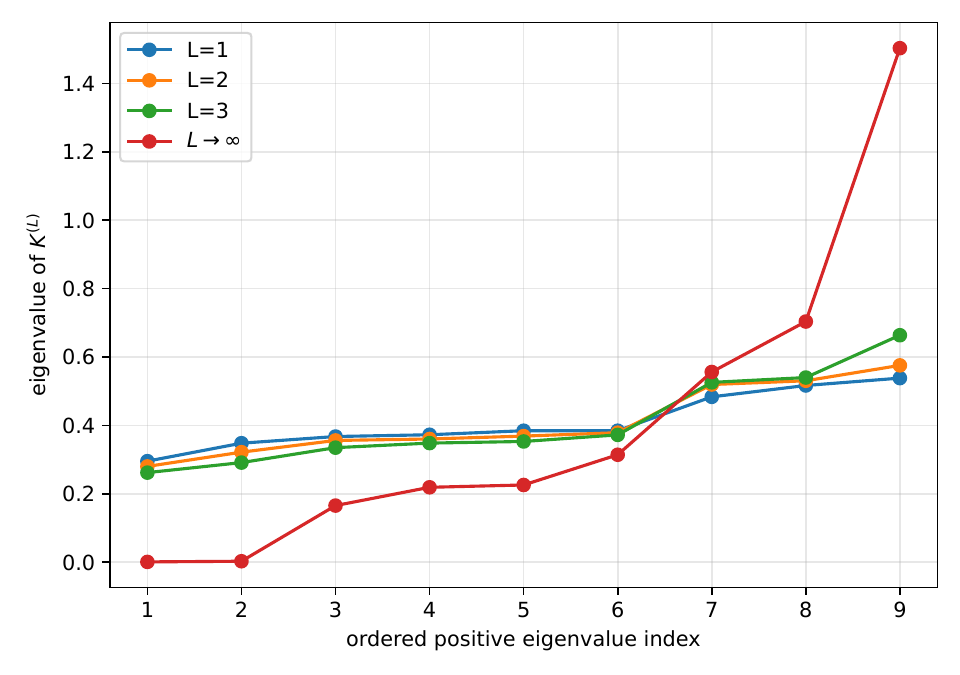}
\caption{Ordered positive eigenvalues of $K^{(L)}(0)$ at the reference background for $L=1,2,3$ and the canonical large-$L$ limit. The zero eigenvalue is omitted. The average separation between the largest three and the remaining six positive eigenvalues is larger at $L=2$ and $L=3$ than at $L=1$, and larger still in the canonical large-$L$ limit. The grouping indicates an ordered diagnostic, not an exact degeneracy.}
\label{fig:ordered-positive-spectra}
\end{figure}

The ordered positive spectra are shown in Fig.~\ref{fig:ordered-positive-spectra}.

Thus, the representation-dependent reweighting makes the SO(5)-resolved structure increasingly visible. It exposes the microscopic redistribution that is hidden in the \(L=1\) quadratic trace and, at the sampled representation levels, gives the ordered \(3+6\) separations quoted above.

In the next section, we compare this representation response with the local one-loop response found in Section 4.

\section{One-loop and SO(5) representation responses}\label{one-loop-and-so5-representation-responses}

In Sections 4 and 5, we studied two different responses along the same physical modulus

\begin{equation}
A_\mu(u).
\end{equation}

For the one-loop comparison, we use the intrinsic fermionic scalar

\begin{equation}
\Delta\mathcal P_F(u)
=
\log|\operatorname{Pf}\mathcal M_F(u)|
-
\log|\operatorname{Pf}\mathcal M_F(0)|.
\end{equation}

For the framed higher-\(L\) response, it is convenient to use the normalized quadratic moment

\begin{equation}
\widehat I_L(u)
:=
\frac{\operatorname{Tr}_{10}[(K^{(L)}(u))^2]}
{[\operatorname{Tr}_{10}K^{(L)}(u)]^2},
\end{equation}

and its root-referenced change

\begin{equation}
\Delta\widehat I_L(u)
=
\widehat I_L(u)-\widehat I_L(0).
\end{equation}

The normalization is stated explicitly because the numerical scatter plot uses \(\Delta\widehat I_L\), not the unnormalized trace moment \(I_L=\operatorname{Tr}_{10}[(K^{(L)})^2]\).

On the same \(23\) accepted backgrounds, the Pearson correlation coefficients of \(\Delta\mathcal P_F\) with \(\Delta\widehat I_2\), \(\Delta\widehat I_3\), \(R_{3+6}(2,u)\), and \(R_{3+6}(3,u)\) are approximately

\begin{equation}
0.99908,
\qquad
0.99943,
\qquad
0.99909,
\qquad
0.99922,
\end{equation}

respectively. A representative pointwise comparison is shown in Appendix E, Fig.~\ref{fig:correlation-one-loop-vs-I2}.

A high correlation between two smooth quantities on a short one-dimensional branch is not, by itself, evidence for a dynamical mechanism. We therefore use this comparison only as a compact description of the sampled profiles. The SO(5)-resolved redistribution is already present at the classical level, while the Pfaffian response is obtained from the fermionic quadratic operator on the full \(N=5\) background. Neither response is taken to generate the other, and the correlation is not needed for the main claim that the physical quotient points are distinguished by an intrinsic one-loop scalar.

In the next section, we discuss the fixed-\(Q_\rho\) description and its continuation in the radial direction.

\section{Scope of the analysis and the radial direction}\label{scope-of-the-analysis-and-the-radial-direction}

\subsection{\texorpdfstring{Meaning of the fixed-\(Q_\rho\) slice}{Meaning of the fixed-Q\_\textbackslash rho slice}}\label{meaning-of-the-fixed-q_rho-slice}

In Sections 3--6, we fixed

\begin{equation}
Q_\rho[A]
=
\operatorname{Tr}(A_\mu A^\mu)-Q_{\rm ref},
\qquad
Q_{\rm ref}=6+2\sqrt3,
\end{equation}

to

\begin{equation}
Q_\rho=9.
\end{equation}

This condition fixes the homogeneous radial normalization of the background and allows us to compare changes in the shape and microscopic noncommutative structure of the matrix configuration.

Near this slice, we introduce

\begin{equation}
s:=Q_\rho
\end{equation}

and consider the homogeneous continuation

\begin{equation}
A_\mu(u;s)
=
b(s)A_\mu(u;9),
\qquad
b(s)^2
=
\frac{s+Q_{\rm ref}}{9+Q_{\rm ref}}.
\end{equation}

Along this continuation, the zero mode corresponding to the physical modulus is preserved. The hard-sector eigenvalues scale as

\begin{equation}
b(s)^2.
\end{equation}

Since \(b(s)^2>0\) in the neighborhood considered here, the hard-sector inertia is unchanged. The detailed scaling relations are given in Appendix F.

The fixed-\(Q_\rho\) stationary landscape also contains several solution families. The physical modulus studied in Sections 3--6 is one representative local branch among them. We do not assume that this branch is unique or dynamically preferred over the other branches. The branch structure and the role of the algebraic special point are discussed in Appendix F.

\subsection{Radial scaling of the one-loop profile}\label{radial-scaling-of-the-one-loop-profile}

Combining the scaling of the one-loop factors under the homogeneous continuation, we obtain the exact factorization

\begin{equation}
\mathcal W_{\rm 1loop}(u,s)
=
b(s)^{92}
\mathcal W_{\rm 1loop}(u,9).
\end{equation}

The origin of the exponent \(92\) is derived in Appendix F.

For each radial slice, we define the relative local one-loop density magnitude by

\begin{equation}
W_{\rm rel}(u;s)
:=
\frac{
\mathcal W_{\rm 1loop}(u,s)
}{
\mathcal W_{\rm 1loop}(0,s)
}.
\end{equation}

The common radial factor then cancels, giving

\begin{equation}
W_{\rm rel}(u;s)
=
W_{\rm rel}(u;9).
\end{equation}

This identity compares density coefficients in the same continuation chart \(u\), transported across the homogeneous radial slices. It does not turn the coefficient itself into a reparametrization-invariant observable.

The intrinsic Pfaffian response has a simple radial statement. Since

\begin{equation}
\operatorname{Pf}\mathcal M_F(u;s)
=
b(s)^{192}\operatorname{Pf}\mathcal M_F(u;9),
\end{equation}

the common radial factor cancels in the root-referenced scalar response,

\begin{equation}
\Delta\mathcal P_F(u;s)
=
\Delta\mathcal P_F(u;9).
\end{equation}

The fixed-section determinant diagnostic also receives only the common shift \(-71\log b(s)\), so its root-referenced profile is unchanged within the same transported Lorentz section. This latter statement remains a fixed-section one and is not promoted to a quotient scalar. The fixed-chart relative density coefficient above is likewise unchanged, with the stated coordinate qualification.

A related statement can be made about radial integration. If the same radial integration prescription can be applied to all values of \(u\) on this factorized homogeneous branch, the radial integration contributes a common factor to the relative \(u\)-profile.

This statement concerns a reduced radial problem. The Lorentzian one-loop calculation in Section 4 defines the finite-dimensional fluctuation integral locally around each real background using an \(i\epsilon\) prescription. The corresponding Lorentzian Gaussian phase is retained, as described in Appendix D.

The full Lorentzian IIB matrix integral requires, in addition, a choice of integration cycle in the complete matrix configuration space. This is a separate global problem; recent Lefschetz-thimble calculations also emphasize that Euclidean and Lorentzian formulations need not be related by a naive contour continuation \cite{ChouNishimuraTripathi:2025}. We therefore distinguish the local finite-dimensional fluctuation prescription, the reduced radial integration, and the global matrix-space integration cycle. Further details are given in Appendix F.

\section{Discussion}\label{discussion}

\subsection{Quantum resolution beyond coarse quadratic geometry}\label{quantum-resolution-beyond-coarse-quadratic-geometry}

The \(N=5\) physical modulus studied in this paper gives a finite-dimensional example in which coarse quadratic geometry and microscopic matrix structure can be separated.

Along the physical modulus, the classical action is common and the \(K\)-spectrum remains common within numerical precision over the analyzed local interval. Nevertheless, the microscopic noncommutative structure changes continuously. At the first nontrivial quantum order, this difference is already visible in the intrinsic Majorana--Weyl Pfaffian magnitude.

This shows that quadratic extent data do not contain all the information relevant to the quantum behavior of a matrix background. Configurations that are indistinguishable at the level of the coarse quadratic spectral data can have different fermionic one-loop responses. The complete one-loop quotient measure is naturally a density one-form on the solution curve; its coefficient depends on the chosen parametrization. The combined hard-determinant expression in the auxiliary Lorentz section is retained only as a fixed-section diagnostic.

Continuous equal-energy matrix-background families with different fluctuation spectra are known in related matrix-model settings \cite{SperlingSteinacker:2018}, and quantum effective dynamics on IIB matrix-model moduli has a long history \cite{Aoki:1998,HartnollLiu:2025}. The bounded distinction here is more specific: on one local, constrained, finite-\(N\) Lorentzian physical branch, the complete \(K\)-spectrum remains common within numerical precision while microscopic noncommutative observables and the intrinsic Pfaffian magnitude vary. In this sense, the one-loop fluctuation response distinguishes a classical degeneracy that is not resolved by the \(K\)-spectrum; it is not being presented as a general discovery of quantum lifting of matrix-model moduli.

\subsection{\texorpdfstring{SO(5)-resolved structure and the \(3+6\) response}{SO(5)-resolved structure and the 3+6 response}}\label{so5-resolved-structure-and-the-36-response}

The SO(5) decomposition gives a more detailed view of the microscopic deformation behind this quantum resolution.

Along the physical modulus, the contributions associated with the \(\mathbf{10}\) and \(\mathbf{14}\) sectors are redistributed while their \(L=1\) combination remains compatible with the common \(K\)-spectrum within numerical precision. The higher-\(L\) construction changes the relative weight of these sectors and makes this hidden redistribution visible in the spectral data. This construction is defined relative to the chosen structural embedding \(\iota\). We have not established independence under changing to a different admissible SO(5) embedding, so the higher-\(L\) results should be regarded as framed structural diagnostics rather than intrinsic observables of \(A_\mu\) alone.

At the reference background, this response appears as a \(3+6\) ordered spectral hierarchy. The average separation between the largest three and the remaining six positive eigenvalues is larger at \(L=2\) and \(L=3\) than at \(L=1\), and larger still in the canonical large-\(L\) limit. These sampled values do not establish monotonicity of \(R_{3+6}(L)\) for every integer \(L\).

The intrinsic Pfaffian response and the framed higher-\(L\) response are highly correlated over the sampled local interval and are therefore consistent with shared sensitivity to the branch deformation. Because the comparison is made on a short one-dimensional branch, we treat this correlation as descriptive rather than as evidence for a common dynamical mechanism or an analytic identity. Whether a more direct analytic relation exists remains an open question.

The \(3+6\) ordered spectral hierarchy should therefore be viewed here as a frame-defined higher-\(L\) diagnostic of microscopic noncommutative structure that is not resolved by the ordinary \(L=1\) quadratic spectral data.

\subsection{Possible implications for dimensional selection}\label{possible-implications-for-dimensional-selection}

A broader question in the Lorentzian IIB matrix model is how a particular number of directions can develop differently from the other matrix directions. This question has been studied through Lorentzian simulations, Pfaffian-based mechanisms, and explicit proposals for an \(SO(3)\times SO(6)\) pattern \cite{Kim:2011cr,NishimuraVernizzi:2000,Nishimura:2019,BrandenbergerPasiecznik:2024}. The present \(N=5\) analysis does not address this dynamical selection itself. It only identifies microscopic quantities that can be tested in such a dynamical setting.

Along the physical modulus, the \(K\)-spectrum remains common within numerical precision over the analyzed local interval, while the SO(5)-resolved microscopic structure changes. The microscopic deformation is distinguished intrinsically by the fermionic Pfaffian response at one-loop order. Relative to the chosen SO(5) structural frame, it is also visible through the higher-\(L\) diagnostic and its \(3+6\) ordered spectral response.

The result is therefore best viewed as a diagnostic statement: information potentially relevant to dimensional dynamics can be carried by microscopic noncommutative structure and one-loop fluctuation response even when the coarse \(K\)-spectrum does not resolve the underlying microscopic deformation. The framed higher-\(L\) construction provides a complementary structural diagnostic, not evidence by itself for dimensional selection, spontaneous symmetry breaking, or compactification.

At the reference background, we find numerically

\begin{equation}
R_{3+6}(1)
<
R_{3+6}(2)
<
R_{3+6}(3),
\end{equation}

and the ordered separation becomes still stronger toward the canonical large-\(L\) limit. This is a structural representation response of the \(N=5\) microscopic data. The representation label \(L\) does not describe a higher-\(N\) dynamical saddle.

A natural next step is therefore to study higher-\(N\) dynamical saddles directly. One can then ask whether a similar SO(5)-resolved sector redistribution occurs dynamically, whether the quantum response follows that redistribution systematically, and whether the three-versus-six spectral separation persists or becomes stronger on those dynamical backgrounds.

The present \(N=5\) result does not establish dimensional selection. It instead identifies microscopic noncommutative structure and one-loop fluctuation response as concrete quantities that can be tested in the higher-\(N\) dynamics.

\section{Conclusion}\label{conclusion}

In the local fixed-\(Q_\rho\) constrained problem studied here at \(N=5\), we found a one-dimensional physical modulus on which the \(K\)-spectrum remains common to the stated numerical precision on the 23 accepted backgrounds spanning the sampled local branch. At the same time, the Majorana--Weyl Pfaffian magnitude varies along the modulus. Thus, backgrounds that are not distinguished by the coarse quadratic spectral data are already distinguished by an intrinsic scalar one-loop response on the gauge/proper-Lorentz quotient. The complete local one-loop measure is a Lorentz-representative-independent density one-form; its coefficient depends on the branch parametrization.

Relative to the chosen SO(5) structural frame, the same microscopic deformation is also visible through a framed higher-\(L\) diagnostic. At the reference background, this diagnostic exhibits a \(3+6\) ordered spectral hierarchy. The close numerical correlation between the intrinsic Pfaffian response and the framed higher-\(L\) response is a branch-local descriptive observation and is not used as evidence for a causal mechanism.

On this local constrained branch, these results show that information absent from the coarse quadratic spectrum is already present in microscopic noncommutative observables and in the intrinsic fermionic one-loop response. Possible relevance to dimensional dynamics remains an interpretation rather than a result of the present calculation. A direct next step is to study higher-\(N\) dynamical saddles and determine whether analogous microscopic structure and fluctuation responses accompany actual dimensional anisotropy.

\section*{Reproducibility and data availability}

A machine-readable reproducibility archive accompanies this manuscript. It contains the \(23\) accepted background matrices \(A_\mu(u_i)\), the continuation coordinates and Lagrange multipliers, pointwise residual and rank diagnostics, the \(K\)-spectra, one-loop pointwise ledgers, higher-\(L\) spectral data, and deterministic replay and validation scripts. The archive keeps the intrinsic Pfaffian response, the fixed-section determinant diagnostic, and the quotient-density coefficient as separate machine-readable quantities. It also includes a proper-Lorentz representative covariance test, figure-generation scripts, the tested numerical environment, acceptance thresholds, basis conventions, and file hashes.

The term \emph{accepted background} in this paper refers to passing these predeclared deterministic floating-point criteria; it does not mean rigorous interval certification. The data and code are archived on Zenodo \cite{MuramatsuData:2026}, DOI: \href{https://doi.org/10.5281/zenodo.22171221}{10.5281/zenodo.22171221}. The main numerical results of Sections 3--6 and Appendices B--E can be replayed from the archived data and scripts.

\clearpage
\section*{Appendices}
\appendix

\section{\texorpdfstring{SO(5) decomposition and trace algebra at general \(L\)}{SO(5) decomposition and trace algebra at general L}}\label{so5-decomposition-and-trace-algebra-at-general-l}

\subsection{\texorpdfstring{SO(5) frame and the \(\mathbf{10}\oplus\mathbf{14}\) decomposition}{SO(5) frame and the \textbackslash mathbf\{10\}\textbackslash oplus\textbackslash mathbf\{14\} decomposition}}\label{so5-frame-and-the-mathbf10oplusmathbf14-decomposition}

In Section 2, we used the decomposition of the traceless matrix space at \(N=5\) into SO(5) representations,

\begin{equation}
\operatorname{End}_0(V_1)
\simeq
\mathbf{10}\oplus\mathbf{14},
\qquad
V_1\simeq\mathbb R^5.
\end{equation}

In this Appendix, we give the representation-theoretic basis of this decomposition and of the general-\(L\) structural representation probe used in Section 5.

We first choose an embedding of SO(5) into the color space,

\begin{equation}
\iota:
\mathfrak{so}(5)\hookrightarrow\mathfrak{su}(5).
\end{equation}

The \(\mathbf{10}\oplus\mathbf{14}\) decomposition below is defined relative to this embedding, or equivalently to the chosen SO(5) structural frame.

For the vector representation \(V_1\simeq\mathbb R^5\),

\begin{equation}
\operatorname{End}(V_1)
\simeq
\mathbf 1
\oplus
\Lambda^2V_1
\oplus
\operatorname{Sym}_0^2V_1,
\end{equation}

with

\begin{equation}
\Lambda^2V_1\simeq\mathbf{10},
\qquad
\operatorname{Sym}_0^2V_1\simeq\mathbf{14}.
\end{equation}

After removing the trace part, we therefore obtain

\begin{equation}
\operatorname{End}_0(V_1)
\simeq
\mathbf{10}\oplus\mathbf{14}.
\end{equation}

Let

\begin{equation}
T_{ab}^{(1)},
\qquad
1\le a<b\le5,
\end{equation}

be the Hermitian generators of the \(\mathbf{10}\) sector, and represent the \(\mathbf{14}\) sector by

\begin{equation}
S_\mu
\in
\operatorname{Sym}_0^2(\mathbb R^5).
\end{equation}

In this frame, the background along the physical modulus can be written as

\begin{equation}
A_\mu(u)
=
\sum_{a<b}
g_\mu^{ab}(u)T_{ab}^{(1)}
+
S_\mu(u).
\end{equation}

Thus,

\begin{equation}
g_\mu^{ab}(u),
\qquad
S_\mu(u),
\end{equation}

are the coefficient data of the \(\mathbf{10}\) and \(\mathbf{14}\) sectors, respectively, for the same \(N=5\) background.

If the color-gauge representative is changed as

\begin{equation}
A_\mu\mapsto UA_\mu U^\dagger,
\end{equation}

the structural frame is transported simultaneously as

\begin{equation}
\iota
\mapsto
\operatorname{Ad}_U\circ\iota.
\end{equation}

The detailed gauge-equivariance of this framed construction is discussed in Appendix E. Here we use the coefficient data defined relative to this structural frame to construct the general-\(L\) probe.

\subsection{\texorpdfstring{Harmonic representations \(V_L\): dimension and Casimir}{Harmonic representations V\_L: dimension and Casimir}}\label{harmonic-representations-v_l-dimension-and-casimir}

For each integer \(L\ge1\), we define

\begin{equation}
V_L
=
\operatorname{Sym}_0^L(\mathbb R^5),
\end{equation}

the symmetric traceless rank-\(L\) representation of SO(5).

The space \(V_L\) can be identified with the space of homogeneous harmonic polynomials of degree \(L\) in five variables. The space of all homogeneous polynomials of degree \(L\) has dimension

\begin{equation}
\binom{L+4}{4}.
\end{equation}

Its trace part is obtained by multiplying \(r^2\) by homogeneous polynomials of degree \(L-2\), whose dimension is

\begin{equation}
\binom{L+2}{4}.
\end{equation}

Therefore,

\begin{equation}
\begin{aligned}
N_L
:=
\dim V_L
&=
\binom{L+4}{4}
-
\binom{L+2}{4}
\\[2mm]
&=
\frac{(L+1)(L+2)(2L+3)}6.
\end{aligned}
\end{equation}

In particular,

\begin{equation}
N_1=5,
\qquad
N_2=14,
\qquad
N_3=30.
\end{equation}

We next determine the quadratic Casimir. In the harmonic-polynomial realization, let

\begin{equation}
J_{ab}
=
x_a\partial_b-x_b\partial_a,
\qquad
T_{ab}=iJ_{ab}.
\end{equation}

Using the Euler operator

\begin{equation}
E=x\cdot\partial
\end{equation}

and the Laplacian \(\Delta\), one finds

\begin{equation}
\sum_{a<b}T_{ab}^2
=
E(E+3)-r^2\Delta.
\end{equation}

On \(V_L\),

\begin{equation}
E=L,
\qquad
\Delta=0,
\end{equation}

and hence

\begin{equation}
\sum_{a<b}
\bigl(T_{ab}^{(L)}\bigr)^2
=
C_2(L)\mathbf1,
\qquad
C_2(L)=L(L+3).
\end{equation}

The trace coefficients used below are completely determined by \(N_L\) and \(C_2(L)\).

Here \(N_L\) is the dimension of the representation. The dynamical backgrounds studied in Sections 3 and 4 remain at \(N=5\). The family \(V_L\) introduced here is a structural representation probe that reads the SO(5)-resolved coefficient data of the same \(N=5\) background through different representations.

\subsection{\texorpdfstring{Equivariant lift of the \(\mathbf{14}\) sector}{Equivariant lift of the \textbackslash mathbf\{14\} sector}}\label{equivariant-lift-of-the-mathbf14-sector}

The \(\mathbf{10}\) sector is lifted directly by coupling the same coefficients \(g_\mu^{ab}\) to the generators \(T_{ab}^{(L)}\) acting on \(V_L\).

For the \(\mathbf{14}\) sector, let

\begin{equation}
S^T=S,
\qquad
\operatorname{tr}_5S=0.
\end{equation}

We construct an operator on \(V_L\) from \(S\) using the map

\begin{equation}
\mathcal E_L(S)
=
\frac16
\sum_{a,b,c=1}^{5}
S_{ab}
\left\{
T_{ac}^{(L)},
T_{bc}^{(L)}
\right\}.
\end{equation}

This map is SO(5)-equivariant and preserves the \(\mathbf{14}\) transformation law of \(S\) on \(V_L\).

Its structure is more transparent in the harmonic-polynomial realization. Define

\begin{equation}
D_S
=
S_{ab}x_a\partial_b,
\qquad
H_S
=
S_{ab}\partial_a\partial_b.
\end{equation}

Then

\begin{equation}
\mathcal E_L(S)
=
\frac13
\left[
(2L+1)D_S-r^2H_S
\right].
\end{equation}

For \(L=1\), the second-derivative term vanishes on degree-one polynomials,

\begin{equation}
H_S=0,
\end{equation}

and therefore

\begin{equation}
\mathcal E_1(S)=S.
\end{equation}

Thus, with this normalization, the \(L=1\) construction reproduces the original \(N=5\) \(\mathbf{14}\) sector.

The canonical lift at general \(L\) is therefore

\begin{equation}
\widetilde A_\mu^{(L)}(u)
=
\sum_{a<b}
g_\mu^{ab}(u)T_{ab}^{(L)}
+
\mathcal E_L\!\left(S_\mu(u)\right).
\end{equation}

Only the SO(5) representation is changed. The coefficient data along the physical modulus,

\begin{equation}
g_\mu^{ab}(u),
\qquad
S_\mu(u),
\end{equation}

remain the same.

\subsection{Exact trace coefficients of the two sectors}\label{exact-trace-coefficients-of-the-two-sectors}

To construct the quadratic extent at general \(L\), we need the trace normalizations of the \(\mathbf{10}\) and \(\mathbf{14}\) sectors. The harmonic representations, Casimir normalization, and fuzzy-\(S^4\) matrix-algebra framework used here are standard \cite{Ramgoolam:2001,Kimura:2002,Sperling:2017}; the purpose is to obtain the normalization appropriate to the framed probe used for the present \(N=5\) data.

\subsubsection{\texorpdfstring{The \(\mathbf{10}\) sector}{The \textbackslash mathbf\{10\} sector}}\label{the-mathbf10-sector}

SO(5) invariance gives

\begin{equation}
\operatorname{Tr}_L
\left(
T_{ab}^{(L)}T_{cd}^{(L)}
\right)
=
\kappa_L
\left(
\delta_{ac}\delta_{bd}
-
\delta_{ad}\delta_{bc}
\right).
\end{equation}

Contracting over the ten independent pairs \(a<b\),

\begin{equation}
\sum_{a<b}
\operatorname{Tr}_L
\left[
\bigl(T_{ab}^{(L)}\bigr)^2
\right]
=
10\kappa_L.
\end{equation}

On the other hand, the Casimir identity gives

\begin{equation}
\sum_{a<b}
\operatorname{Tr}_L
\left[
\bigl(T_{ab}^{(L)}\bigr)^2
\right]
=
N_LC_2(L).
\end{equation}

Thus,

\begin{equation}
\kappa_L
=
\frac{N_LC_2(L)}{10}.
\end{equation}

\subsubsection{\texorpdfstring{The \(\mathbf{10}\)--\(\mathbf{14}\) cross term}{The \textbackslash mathbf\{10\}--\textbackslash mathbf\{14\} cross term}}\label{the-mathbf10mathbf14-cross-term}

The generators \(T_{ab}^{(L)}\) transform in the irreducible \(\mathbf{10}\), whereas \(\mathcal E_L(S)\) transforms in the irreducible \(\mathbf{14}\). The trace pairing

\begin{equation}
(X,Y)\mapsto\operatorname{Tr}_L(XY)
\end{equation}

is SO(5)-invariant. Schur orthogonality between the two inequivalent irreducible representations therefore gives

\begin{equation}
\operatorname{Tr}_L
\left[
T_{ab}^{(L)}\mathcal E_L(S)
\right]
=
0.
\end{equation}

The quadratic trace consequently separates into the sum of the two sector contributions.

\subsubsection{\texorpdfstring{The \(\mathbf{14}\) sector}{The \textbackslash mathbf\{14\} sector}}\label{the-mathbf14-sector}

SO(5) invariance similarly implies

\begin{equation}
\operatorname{Tr}_L
\left[
\mathcal E_L(S)\mathcal E_L(S')
\right]
=
\beta_L\,
\operatorname{tr}_5(SS').
\end{equation}

Let \(S^A\), \(A=1,\ldots,14\), be an orthonormal basis of

\begin{equation}
\operatorname{Sym}_0^2(\mathbb R^5),
\end{equation}

and define

\begin{equation}
Z_L
=
\sum_{A=1}^{14}
\mathcal E_L(S^A)^2.
\end{equation}

Since \(Z_L\) is SO(5)-invariant, it acts as a scalar operator on \(V_L\). Moreover, because \(\mathcal E_L\) is quadratic in the generators, \(Z_L\) is a central element of filtered degree at most four in the universal enveloping algebra. Its central character on the highest-weight family \(L\omega_1\) is therefore a polynomial in \(L\) of degree at most four. Weyl invariance of the shifted highest weight gives the reflection symmetry

\begin{equation}
L\longmapsto -L-3.
\end{equation}

Any polynomial of degree at most four with this symmetry is a polynomial of degree at most two in

\begin{equation}
C_2(L)=L(L+3).
\end{equation}

Equivalently, the possible quartic central character is already included through the \(C_2(L)^2\) term on this one-parameter highest-weight family. More explicitly, after the standard Weyl shift the one-parameter family is identified by the reflection \(L\mapsto-L-3\). Any polynomial in \(L\) of degree at most four that is invariant under this reflection is therefore a polynomial of degree at most two in the basic invariant \(L(L+3)=C_2(L)\). There is no constant term because, for the trivial representation \(L=0\), both the generators and \(\mathcal E_L\) vanish. Hence

\begin{equation}
\frac{\beta_L}{N_L}
=
a\,C_2(L)^2+b\,C_2(L).
\end{equation}

This makes calibration at the following two low representations an exact determination rather than an interpolation.

For \(L=1\),

\begin{equation}
\mathcal E_1(S)=S,
\end{equation}

and therefore

\begin{equation}
\beta_1=1.
\end{equation}

For \(L=2\), identifying \(V_2\) with the space of symmetric traceless matrices \(R\), the exact action is

\begin{equation}
\mathcal E_2(S)R
=
\frac53(SR+RS)
-
\frac23\operatorname{tr}_5(SR)\mathbf1.
\end{equation}

Evaluating the trace with a unit-norm diagonal \(S\), the ten off-diagonal symmetric modes contribute \(25/3\), while the four-dimensional diagonal traceless subspace contributes \(20/3\). Thus,

\begin{equation}
\beta_2
=
\frac{25}{3}
+
\frac{20}{3}
=
15.
\end{equation}

Using

\begin{equation}
C_2(1)=4,
\qquad
N_1=5,
\end{equation}

and

\begin{equation}
C_2(2)=10,
\qquad
N_2=14,
\end{equation}

we obtain

\begin{equation}
16a+4b=\frac15,
\end{equation}

\begin{equation}
100a+10b=\frac{15}{14}.
\end{equation}

Solving these equations gives

\begin{equation}
a=\frac1{105},
\qquad
b=\frac1{84}.
\end{equation}

Therefore,

\begin{equation}
\beta_L
=
\frac{
N_LC_2(L)[4C_2(L)+5]
}{420}.
\end{equation}

The relative quadratic-trace weight of the \(\mathbf{14}\) sector with respect to the \(\mathbf{10}\) sector is consequently

\begin{equation}
\rho_L
:=
\frac{\beta_L}{\kappa_L}
=
\frac{4C_2(L)+5}{42}
=
\frac{4L(L+3)+5}{42}.
\end{equation}

In particular,

\begin{equation}
\rho_1=\frac12,
\qquad
\rho_2=\frac{15}{14},
\qquad
\rho_3=\frac{11}{6}.
\end{equation}

Furthermore,

\begin{equation}
\rho_{L+1}-\rho_L
=
\frac{4(L+2)}{21}>0.
\end{equation}

Thus, the relative trace weight \(\rho_L\) increases exactly with \(L\).

\subsection{Normalized extent operator and sector redistribution}\label{normalized-extent-operator-and-sector-redistribution}

We define the quadratic contributions of the two SO(5) sectors by

\begin{equation}
\mathsf X_{\mu\nu}(u)
=
\sum_{a<b}
g_\mu^{ab}(u)g_\nu^{ab}(u),
\end{equation}

and

\begin{equation}
\mathsf Y_{\mu\nu}(u)
=
\operatorname{tr}_5
\left[
S_\mu(u)S_\nu(u)
\right].
\end{equation}

Using the trace identities derived above, the quadratic tensor before normalization is

\begin{equation}
G_{\mu\nu}^{(L,\mathrm{raw})}(u)
=
\frac{C_2(L)}{10}
\left[
\mathsf X_{\mu\nu}(u)
+
\rho_L\mathsf Y_{\mu\nu}(u)
\right].
\end{equation}

We also define the Lorentzian contractions

\begin{equation}
x(u)
=
\eta^{\mu\nu}\mathsf X_{\mu\nu}(u),
\qquad
y(u)
=
\eta^{\mu\nu}\mathsf Y_{\mu\nu}(u).
\end{equation}

These are contractions with the Lorentz metric. They are not positive-definite norms.

For the original fixed-\(Q_\rho\) background,

\begin{equation}
\operatorname{Tr}(A_\mu A^\mu)
=
9+Q_{\rm ref}
=
15+2\sqrt3.
\end{equation}

We set

\begin{equation}
D_0:=15+2\sqrt3.
\end{equation}

To align the overall quadratic scale of the structural probe across representations, we introduce a single amplitude \(\alpha_L(u)\) such that the per-color quadratic contraction remains

\begin{equation}
\frac{D_0}{5}.
\end{equation}

This gives

\begin{equation}
\alpha_L(u)^2
=
\frac{
2D_0
}{
C_2(L)[x(u)+\rho_Ly(u)]
}.
\end{equation}

After this normalization, the higher-\(L\) Lorentzian extent operator is

\begin{equation}
\bigl(K^{(L)}(u)\bigr)^\mu{}_{\nu}
=
\frac{
D_0
}{
5[x(u)+\rho_Ly(u)]
}
\eta^{\mu\lambda}
\left[
\mathsf X_{\lambda\nu}(u)
+
\rho_L\mathsf Y_{\lambda\nu}(u)
\right].
\end{equation}

The overall factor \(C_2(L)\) has completely canceled. The representation dependence of the normalized higher-\(L\) Lorentzian extent operator is therefore produced by the relative reweighting of the two SO(5) sectors through

\begin{equation}
\rho_L,
\end{equation}

rather than by the overall growth of the Casimir itself.

For \(L=1\),

\begin{equation}
\rho_1=\frac12,
\end{equation}

and

\begin{equation}
G_{\mu\nu}^{(1)}
=
\frac25
\left[
\mathsf X_{\mu\nu}
+
\frac12\mathsf Y_{\mu\nu}
\right].
\end{equation}

Thus, the ordinary \(N=5\) quadratic trace does not retain \(\mathsf X_{\mu\nu}\) and \(\mathsf Y_{\mu\nu}\) separately. It reads the particular combination

\begin{equation}
\mathsf X+\frac12\mathsf Y.
\end{equation}

This cancellation is also seen directly in the tangent data at the reference point \(u=0\). We find

\begin{equation}
\left.\frac{dx}{du}\right|_{u=0}
=
-0.02672,
\end{equation}

and

\begin{equation}
\left.\frac{dy}{du}\right|_{u=0}
=
0.05344,
\end{equation}

whereas

\begin{equation}
\left.
\frac{d}{du}
\left(
x+\frac12y
\right)
\right|_{u=0}
=
3.47\times10^{-18}.
\end{equation}

Thus, the contributions of the two sectors vary separately, while their first-order variation cancels at the \(L=1\) trace weight.

To examine this structure along the branch, we use

\begin{equation}
I_L(u)
=
\operatorname{Tr}_{10}
\left[
\bigl(K^{(L)}(u)\bigr)^2
\right].
\end{equation}

The pointwise difference from the reference background \(u=0\) is

\begin{equation}
\Delta I_L(u)
=
I_L(u)-I_L(0),
\end{equation}

while the sampled branch variation over the same \(23\) accepted backgrounds is defined by

\begin{equation}
\operatorname{span}_u I_L
=
\max_u I_L(u)-\min_u I_L(u).
\end{equation}

On the analyzed local branch, we obtain

\begin{longtable}[]{@{}
  >{\raggedleft\arraybackslash}p{(\columnwidth - 6\tabcolsep) * \real{0.2500}}
  >{\raggedleft\arraybackslash}p{(\columnwidth - 6\tabcolsep) * \real{0.2500}}
  >{\raggedleft\arraybackslash}p{(\columnwidth - 6\tabcolsep) * \real{0.2500}}
  >{\raggedleft\arraybackslash}p{(\columnwidth - 6\tabcolsep) * \real{0.2500}}@{}}
\toprule\noalign{}
\begin{minipage}[b]{\linewidth}\raggedleft
\(L\)
\end{minipage} & \begin{minipage}[b]{\linewidth}\raggedleft
maximum change in the \(K^{(L)}\)-spectrum
\end{minipage} & \begin{minipage}[b]{\linewidth}\raggedleft
\(\operatorname{span}_u I_L\)
\end{minipage} & \begin{minipage}[b]{\linewidth}\raggedleft
\(\left.dI_L/du\right|_{u=0}\)
\end{minipage} \\
\midrule\noalign{}
\endhead
\bottomrule\noalign{}
\endlastfoot
1 & \(1.78\times10^{-15}\) & numerical roundoff level & \(-1.13\times10^{-16}\) \\
2 & \(5.07\times10^{-4}\) & \(1.83\times10^{-4}\) & \(-1.69\times10^{-3}\) \\
3 & \(1.04\times10^{-3}\) & \(5.57\times10^{-4}\) & \(-5.12\times10^{-3}\) \\
\end{longtable}

Thus, at \(L=1\), no spectral variation is resolved beyond numerical precision, whereas for \(L=2\) and \(L=3\) the same physical modulus produces a clear spectral response.

The same point can be expressed using a continuous response parameter \(\varrho\). Formally, define

\begin{equation}
K(\varrho,u)
=
\frac{
D_0
}{
5[x(u)+\varrho y(u)]
}
\eta
\left[
\mathsf X(u)+\varrho\mathsf Y(u)
\right].
\end{equation}

Then

\begin{equation}
\left.
\partial_\varrho\partial_u
\operatorname{Tr}_{10}
\left[
K(\varrho,u)^2
\right]
\right|_{\varrho=1/2,u=0}
=
-0.002166.
\end{equation}

Therefore, once the relative weighting is varied away from the \(L=1\) trace weight \(\varrho=1/2\), sensitivity to the sector redistribution hidden at \(L=1\) appears immediately.

\subsection{\texorpdfstring{The \(3+6\) ordered spectral hierarchy at large \(L\)}{The 3+6 ordered spectral hierarchy at large L}}\label{the-36-ordered-spectral-hierarchy-at-large-l}

At the reference background \(u=0\), \(K^{(L)}\) has one numerical zero eigenvalue and nine positive eigenvalues. We order the positive eigenvalues as

\begin{equation}
0<
\lambda_1^{(L)}
\le\cdots\le
\lambda_9^{(L)}.
\end{equation}

To measure the average separation between the largest three and the remaining six, we define

\begin{equation}
R_{3+6}(L,u)
=
\frac{
\bigl(
\lambda_7^{(L)}
+\lambda_8^{(L)}
+\lambda_9^{(L)}
\bigr)/3
}{
\bigl(
\lambda_1^{(L)}
+\cdots+
\lambda_6^{(L)}
\bigr)/6
}.
\end{equation}

For the reference background, we use the shorthand

\begin{equation}
R_{3+6}(L)
:=
R_{3+6}(L,0).
\end{equation}

The numerical values are

\begin{longtable}[]{@{}
  >{\raggedleft\arraybackslash}p{(\columnwidth - 8\tabcolsep) * \real{0.2000}}
  >{\raggedleft\arraybackslash}p{(\columnwidth - 8\tabcolsep) * \real{0.2000}}
  >{\raggedleft\arraybackslash}p{(\columnwidth - 8\tabcolsep) * \real{0.2000}}
  >{\raggedleft\arraybackslash}p{(\columnwidth - 8\tabcolsep) * \real{0.2000}}
  >{\raggedleft\arraybackslash}p{(\columnwidth - 8\tabcolsep) * \real{0.2000}}@{}}
\toprule\noalign{}
\begin{minipage}[b]{\linewidth}\raggedleft
\(L\)
\end{minipage} & \begin{minipage}[b]{\linewidth}\raggedleft
\(N_L\)
\end{minipage} & \begin{minipage}[b]{\linewidth}\raggedleft
\(\rho_L\)
\end{minipage} & \begin{minipage}[b]{\linewidth}\raggedleft
\(R_{3+6}(L)\)
\end{minipage} & \begin{minipage}[b]{\linewidth}\raggedleft
\(\lambda_7^{(L)}/\lambda_6^{(L)}\)
\end{minipage} \\
\midrule\noalign{}
\endhead
\bottomrule\noalign{}
\endlastfoot
1 & 5 & \(1/2\) & 1.429 & 1.257 \\
2 & 14 & \(15/14\) & 1.574 & 1.372 \\
3 & 30 & \(11/6\) & 1.763 & 1.412 \\
\(L\to\infty\) & --- & \(\infty\) & 5.946 & 1.770 \\
\end{longtable}

On the same \(23\) accepted backgrounds, the sampled variations are

\begin{equation}
\operatorname{span}_uR_{3+6}(1,u)
=
4.9\times10^{-15},
\end{equation}

\begin{equation}
\operatorname{span}_uR_{3+6}(2,u)
=
7.30\times10^{-4},
\end{equation}

and

\begin{equation}
\operatorname{span}_uR_{3+6}(3,u)
=
1.82\times10^{-3}.
\end{equation}

The increase of \(\rho_L\) is an exact representation-theoretic result. In contrast, the sequence of \(R_{3+6}\) values in the table is a numerical spectral result for the reference background.

Since

\begin{equation}
\rho_L\longrightarrow\infty
\qquad
(L\to\infty),
\end{equation}

we obtain

\begin{equation}
K^{(L)}
\longrightarrow
K_\infty
=
\frac{D_0}{5y}
\eta\mathsf Y.
\end{equation}

Thus, in the canonical large-\(L\) limit of this structural representation probe, the \(\mathbf{14}\) sector dominates the normalized spectral shape.

The nine positive eigenvalues in this limit do not form a uniform \(3+6\) degeneracy. More precisely, their internal structure is approximately

\begin{equation}
3_{\rm large}
+
4_{\rm intermediate}
+
2_{\rm very\ small}.
\end{equation}

We therefore use \(R_{3+6}\) as an ordered spectral diagnostic that measures the average separation between the largest three and the remaining six eigenvalues. It does not denote an exact \(3+6\) degeneracy.

\subsection{Kernel direction and SO(5) structure}\label{kernel-direction-and-so5-structure}

We finally examine the relation between the one-dimensional kernel present at the reference background and the representation lift.

From the coefficient data, define

\begin{equation}
C(n)
=
\left(
\left\{
\sum_\mu n^\mu g_\mu^{ab}
\right\}_{a<b},
\,
\sum_\mu n^\mu S_\mu
\right).
\end{equation}

Since \(\mathsf X\) and \(\mathsf Y\) are the quadratic forms associated with the two sets of coefficient data,

\begin{equation}
\ker C
=
\ker\mathsf X
\cap
\ker\mathsf Y
=
\ker
\left(
\mathsf X+\varrho\mathsf Y
\right),
\qquad
\varrho>0.
\end{equation}

Therefore, a kernel already present in the coefficient map is preserved when the relative weight \(\varrho\), and hence the representation level \(L\), is changed.

Numerically, for the reference background considered here, we find

\begin{equation}
\operatorname{rank}C=9,
\qquad
\operatorname{nullity}C=1.
\end{equation}

The two smallest singular values relevant for this separation are

\begin{equation}
\sigma_{\min}
=
4.0\times10^{-15},
\end{equation}

and

\begin{equation}
\sigma_{\rm next}
=
0.875.
\end{equation}

They indicate an isolated one-dimensional kernel within numerical precision.

For the sole purpose of normalizing the ten-component vector \(n^\mu\), we use the auxiliary positive-definite Euclidean norm on its component space. This normalization is not part of the Lorentzian dynamics. With \(n\) normalized to unit Euclidean norm, its Lorentzian contraction is

\begin{equation}
n^\mu\eta_{\mu\nu}n^\nu
=
-0.980.
\end{equation}

Thus, for this background, the spacetime direction corresponding to the numerically identified zero eigenvalue is timelike.

This timelike kernel is a property of the coefficient data of the background. It is not forced by SO(5) representation theory itself. For example, if the coefficients of the \(\mathbf{10}\) sector are assigned independently to the ten spacetime directions as

\begin{equation}
g_\mu^{I}=\delta_\mu^{I},
\qquad
S_\mu=0,
\end{equation}

the coefficient map has rank \(10\) and no kernel.

On the other hand, the kernel present in the background studied here is preserved for every \(\varrho>0\). The same kernel direction is therefore retained by the general-\(L\) structural representation probe.

The results of this Appendix show that the higher-\(L\) response in Section 5 is obtained by reading the \(\mathbf{10}\) and \(\mathbf{14}\) coefficient data of the same \(N=5\) physical background with the exact representation-dependent relative trace weights \(\rho_L\). The sector redistribution hidden by the \(L=1\) trace becomes visible as a spectral response for \(L>1\). At the reference background, the same representation-dependent reweighting also makes the \(3+6\) ordered spectral hierarchy more pronounced.

\section{Constrained quotient and physical modulus}\label{constrained-quotient-and-physical-modulus}

\subsection{\texorpdfstring{Physical tangent space at \(N=5\)}{Physical tangent space at N=5}}\label{physical-tangent-space-at-n5}

In Section 3, we showed that the fixed-\(Q_\rho\) constrained problem contains a one-dimensional physical modulus. In this Appendix, we first construct the physical tangent space by removing the redundant directions associated with symmetries and the fixed normalization condition. We then show that the single zero mode in this space extends to an actual continuous family of solutions.

At \(N=5\), one traceless Hermitian matrix has

\begin{equation}
5^2-1=24
\end{equation}

real degrees of freedom. The traceless configuration space of the ten bosonic matrices therefore has dimension

\begin{equation}
\dim E_{\rm bos}
=
10(5^2-1)
=
240.
\end{equation}

Infinitesimal color-gauge transformations are generated by

\begin{equation}
\delta_\xi A_\mu
=
i[\xi,A_\mu],
\qquad
\xi\in\mathfrak{su}(5).
\end{equation}

For the solution family analyzed here, the gauge orbit has \(24\) independent tangent directions. The local gauge quotient is therefore \(216\)-dimensional.

We next fix

\begin{equation}
Q_\rho[A]
=
\operatorname{Tr}(A_\mu A^\mu)-Q_{\rm ref},
\qquad
Q_{\rm ref}=6+2\sqrt3,
\end{equation}

to

\begin{equation}
Q_\rho=9.
\end{equation}

In the neighborhood of the backgrounds analyzed here, this gives one regular level-set condition. The tangent space to the fixed-\(Q_\rho\) gauge quotient is therefore \(215\)-dimensional.

We then remove directions that represent the same configuration in different Lorentz frames. An infinitesimal proper Lorentz transformation acts as

\begin{equation}
\delta_\omega A_\mu
=
\omega_\mu{}^\nu A_\nu,
\qquad
\omega_{\mu\nu}=-\omega_{\nu\mu}.
\end{equation}

On the analyzed branch, the Lorentz-frame tangent map has rank \(45\) after the gauge directions and the \(Q_\rho\)-normal direction have been removed. This number is the actual projected tangent rank on the analyzed background family; it is not inferred merely from the dimension of the ten-dimensional Lorentz group.

The dimension of the tangent space describing independent physical fluctuations is therefore

\begin{equation}
240-24-1-45=170.
\end{equation}

This \(170\)-dimensional space is not the dimension of the physical modulus. It is the physical fluctuation space that remains after removing the redundancies associated with the symmetries and the normalization condition. The number of flat directions within this space is determined by the constrained Hessian.

\subsection{Constrained Hessian and the unique physical zero mode}\label{constrained-hessian-and-the-unique-physical-zero-mode}

We define the constrained Lagrange function by

\begin{equation}
\mathcal L_Q[A,\lambda_Q]
=
S_B[A]
-
\lambda_Q
\bigl(Q_\rho[A]-9\bigr).
\end{equation}

Its stationary equation is

\begin{equation}
[A^\nu,[A_\nu,A_\mu]]
=
2\lambda_QA_\mu.
\end{equation}

We denote by \(H_{\rm phys}\) the second variation of \(\mathcal L_Q\) restricted to the \(170\)-dimensional physical tangent space constructed in the previous subsection.

Along the solution family analyzed here, only one of these \(170\) directions is a zero mode. The remaining \(169\) directions have nonzero quadratic curvature. Their inertia is

\begin{equation}
(95,74,1),
\end{equation}

with \(95\) positive directions, \(74\) negative directions, and one zero direction. This decomposition remains unchanged along the analyzed branch.

Locally, the physical tangent space therefore separates into

\begin{equation}
\text{one zero-mode direction}
\quad+\quad
\text{169 transverse directions}.
\end{equation}

In Section 4, we denote the restriction of \(H_{\rm phys}\) to this \(169\)-dimensional transverse subspace by

\begin{equation}
H_{\rm hard}.
\end{equation}

Here ``hard'' means only that these directions have nonzero quadratic curvature.

The existence of a zero direction in the Hessian does not by itself show that it corresponds to a finite classical deformation. To establish this, the full constrained equations must be followed nonlinearly away from the reference background.

\subsection{Nonlinear continuation of the zero mode}\label{nonlinear-continuation-of-the-zero-mode}

Starting from the zero mode of the physical Hessian, we numerically continued the full constrained equations in both directions.

We choose the reference background at

\begin{equation}
u=0
\end{equation}

and use \(u\) as a local coordinate along the solution family.

For the continuation, we used a pseudo-arclength method. At each step, a prediction was made along the zero-mode direction, followed by a correction that simultaneously imposed the constrained equations and

\begin{equation}
Q_\rho=9.
\end{equation}

This procedure produced a set of \(23\) stationary backgrounds spanning both sides of \(u=0\). We analyzed the local interval

\begin{equation}
-0.05\lesssim u\lesssim0.05.
\end{equation}

Over this interval, the residual of the constrained equations remains at the level of \(10^{-16}\), while the variation of \(Q_\rho\) remains at the level of \(10^{-14}\).

The physical zero-mode count remains one throughout the interval, and the other \(169\) eigenvalues remain separated from zero. The smallest magnitude among the transverse eigenvalues is approximately

\begin{equation}
4.2\times10^{-2}.
\end{equation}

The tangent to the numerically obtained solution curve also agrees with the zero direction of the physical Hessian.

Thus, the zero mode is not only a flat direction of the quadratic approximation. It is tangent to a one-dimensional local constrained critical manifold,

\begin{equation}
u\longmapsto A_\mu(u).
\end{equation}

This local continuous family is the physical modulus used throughout the main text.

\subsection{\texorpdfstring{Classical action and the \(K\)-spectrum}{Classical action and the K-spectrum}}\label{classical-action-and-the-k-spectrum}

At a constrained stationary point,

\begin{equation}
dS_B
=
\lambda_Q\,dQ_\rho.
\end{equation}

Therefore, along the fixed-\(Q_\rho\) family,

\begin{equation}
dQ_\rho=0
\end{equation}

implies

\begin{equation}
dS_B=0.
\end{equation}

For an exact constrained solution family, the classical action is therefore constant by the stationary identity. On the accepted numerical backgrounds, the corresponding variation remains at the level of \(10^{-14}\), consistent with the quoted residual accuracy.

We next consider the Lorentzian extent operator introduced in Section 2,

\begin{equation}
K^\mu{}_\nu
=
\eta^{\mu\rho}G_{\rho\nu}.
\end{equation}

We denote its eigenvalues by

\begin{equation}
\lambda_i^K(A),
\qquad
i=1,\ldots,10,
\end{equation}

and write the \(K\)-spectrum as

\begin{equation}
\mathcal S_K(A)
=
\{\lambda_i^K(A)\}_{i=1}^{10}.
\end{equation}

Comparing these eigenvalues over the \(23\) analyzed backgrounds gives

\begin{equation}
\max_{u,i}
\left|
\lambda_i^K(A(u))
-
\lambda_i^K(A(0))
\right|
\simeq
2.0\times10^{-15}.
\end{equation}

\begin{figure}[t]
\centering
\includegraphics[width=0.82\textwidth]{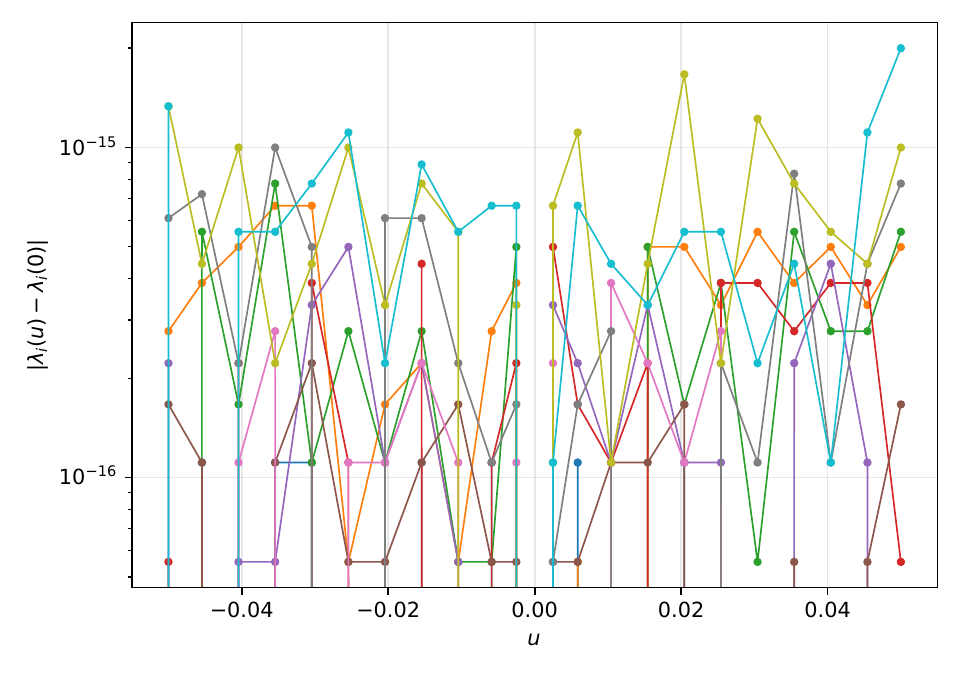}
\caption{Numerical deviations of the ten eigenvalues of $K^\mu{}_{\nu}$ from the reference background along the physical modulus. The maximum absolute deviation over the $23$ accepted backgrounds is about $2.0\times 10^{-15}$, supporting the statement that the $K$-spectrum remains common within numerical precision over the analyzed local interval.}
\label{fig:k-spectrum-deviations}
\end{figure}

Independent comparisons using trace powers and the coefficients of the characteristic polynomial show the same level of constancy. The numerical deviations of the ten eigenvalues from the reference background are shown in Fig.~\ref{fig:k-spectrum-deviations}.

Thus, the result established here is that

\begin{equation}
\text{the \(K\)-spectrum is constant within numerical precision
over the analyzed local interval.}
\end{equation}

This statement does not mean that each component of

\begin{equation}
G_{\mu\nu}
\end{equation}

is constant in a particular Lorentz frame. In a fixed numerical frame, the components of \(G_{\mu\nu}\) vary at the level of \(10^{-4}\). What remains common within numerical precision is the Lorentz-frame-independent \(K\)-spectrum.

\subsection{Variation of the microscopic noncommutative structure}\label{variation-of-the-microscopic-noncommutative-structure-1}

A common \(K\)-spectrum within numerical precision does not imply that all information contained in the original matrix configurations is the same.

To examine the microscopic difference, we use

\begin{equation}
Q_2[A]
:=
\operatorname{Tr}(A_\mu A^\mu)
=
Q_\rho[A]+Q_{\rm ref}
\end{equation}

to normalize the homogeneous scale. Here \(Q_2[A]\) is a Lorentzian quadratic contraction, not a positive-definite norm.

We then consider the sixth-order invariant

\begin{equation}
I_{6,\triangle}[A]
=
\frac{1}{NQ_2[A]^3}
\operatorname{Tr}
\left(
A_\mu A_\nu A_\rho
A^\mu A^\nu A^\rho
\right),
\end{equation}

which is sensitive to the ordering of the noncommuting matrices.

At the reference background,

\begin{equation}
I_{6,\triangle}(0)
=
2.89\times10^{-4}.
\end{equation}

This quantity changes along the physical modulus. Over the analyzed interval, its variation reaches approximately

\begin{equation}
3.6\times10^{-6}.
\end{equation}

A similar variation is found in independent gauge- and Lorentz-invariant diagnostics constructed from the commutators.

The observed microscopic variation therefore does not depend on the choice of a single higher-order invariant.

Because these quantities are invariant under color-gauge transformations and Lorentz transformations, backgrounds at different values of \(u\) are not simply different gauge representatives or Lorentz-frame representatives of the same matrix configuration.

The local solution family therefore satisfies

\begin{equation}
Q_\rho=\mathrm{const.},
\qquad
S_B=\mathrm{const.},
\end{equation}

while its \(K\)-spectrum remains common within numerical precision and its higher-order noncommutative invariants vary continuously.

Thus, the physical modulus exhibits continuously different microscopic noncommutative structures behind the same classical action and coarse quadratic spectral data within the numerical accuracy established above.

In Section 4, we keep this one-dimensional direction as the collective coordinate \(u\) and treat the \(169\) physical transverse modes as Gaussian fluctuations. The corresponding finite-dimensional one-loop operators are described in Appendix C.

\section{Finite-dimensional one-loop fluctuation operators}\label{finite-dimensional-one-loop-fluctuation-operators}

\subsection{\texorpdfstring{Quadratic fluctuations around the \(N=5\) backgrounds}{Quadratic fluctuations around the N=5 backgrounds}}\label{quadratic-fluctuations-around-the-n5-backgrounds}

In Section 4, we integrated the quadratic fluctuations around each classical background on the one-dimensional physical modulus,

\begin{equation}
A_\mu=A_\mu(u).
\end{equation}

In this Appendix, we summarize the bosonic, ghost, and Majorana--Weyl fermionic operators used in this calculation.

Around each background, we write

\begin{equation}
A_\mu=A_\mu(u)+a_\mu,
\end{equation}

and consider fermionic fluctuations around the background value

\begin{equation}
\psi=0.
\end{equation}

We define the adjoint derivative associated with the background \(A_\mu(u)\) by

\begin{equation}
D_\mu X
=
i[A_\mu(u),X].
\end{equation}

At \(N=5\), all the quadratic operators can be constructed directly as finite-dimensional matrices. The one direction along the physical modulus found in Section 3 is retained as the collective coordinate \(u\). The \(169\) directions transverse to it are treated as bosonic quadratic fluctuations.

The dimensions of the one-loop Gaussian sectors are therefore

\begin{equation}
169_{\rm\ bosonic},
\qquad
24_{\rm\ ghost},
\qquad
384_{\rm\ fermionic}.
\end{equation}

The collective-coordinate measure associated with the physical modulus is introduced in Appendix D.

\subsection{Physical bosonic hard operator}\label{physical-bosonic-hard-operator}

As shown in Appendix B, the physical tangent space obtained after removing the color-gauge, fixed-\(Q_\rho\), and Lorentz-frame directions is \(170\)-dimensional. It contains one zero direction corresponding to the physical modulus. The remaining \(169\) directions have nonzero quadratic curvature.

We define the constrained Lagrange function by

\begin{equation}
\mathcal L_Q[A,\lambda_Q]
=
S_B[A]
-
\lambda_Q
\bigl(Q_\rho[A]-9\bigr),
\end{equation}

and denote by

\begin{equation}
H_{\rm phys}(u)
\end{equation}

its second variation restricted to this \(170\)-dimensional physical tangent space.

To define the normalization and orthogonal decomposition used below, we use the auxiliary positive-definite inner product on the real coefficient space of the bosonic matrix components introduced in Section 4.2,

\begin{equation}
\langle a,b\rangle_{\rm coeff}
=
\delta^{\mu\nu}\operatorname{Tr}(a_\mu b_\nu).
\end{equation}

This inner product is used only to choose finite-dimensional normalized bases and orthogonal projections in coefficient space. It is an auxiliary structure and is distinct from the Lorentzian metric \(\eta_{\mu\nu}\) entering the action and the constraint.

Let \(t(u)\) be the normalized tangent to the physical modulus, and let \(U_h(u)\) be an orthonormal basis of its orthogonal complement. Then

\begin{equation}
U_h^{\mathsf T}U_h
=
I_{169},
\qquad
U_h^{\mathsf T}t
=
0.
\end{equation}

The hard bosonic operator used in Section 4 is defined by

\begin{equation}
H_{\rm hard}(u)
=
U_h(u)^{\mathsf T}
H_{\rm phys}(u)
U_h(u).
\end{equation}

Here \(t(u)\) is determined from the tangent to the nonlinear solution curve. Thus, \(H_{\rm hard}(u)\) is the Hessian for the \(169\) independent physical fluctuations transverse to the physical modulus.

On the analyzed local branch, all \(169\) eigenvalues are nonzero. Their inertia consists of \(95\) positive and \(74\) negative directions, and this sign structure is preserved along the branch.

The hard-sector spectrum gives the bosonic magnitude factor

\begin{equation}
\frac{1}{
\sqrt{
|\det H_{\rm hard}(u)|
}
}.
\end{equation}

The Lorentzian stationary-phase factor and the local \(i\epsilon\) prescription are discussed in Appendix D.

\subsection{Background gauge and the Faddeev--Popov operator}\label{background-gauge-and-the-faddeevpopov-operator}

We treat the color-gauge redundancy at quadratic order using the background-gauge condition

\begin{equation}
D^\mu a_\mu=0.
\end{equation}

The corresponding gauge-fixing term is

\begin{equation}
S_{\rm gf}
=
\frac{1}{2g^2}
\operatorname{Tr}
\left(
D^\mu a_\mu
\right)^2.
\end{equation}

Under an infinitesimal color-gauge transformation,

\begin{equation}
\delta_\xi a_\mu
=
D_\mu\xi,
\end{equation}

the gauge condition changes as

\begin{equation}
\delta_\xi(D^\mu a_\mu)
=
D^\mu D_\mu\xi.
\end{equation}

With the sign convention used in this paper, we therefore define the Faddeev--Popov operator by

\begin{equation}
\Delta_{\rm FP}(u)
=
-D_\mu D^\mu.
\end{equation}

The quadratic ghost action is

\begin{equation}
S_{\rm gh}^{(2)}
=
\frac{1}{g^2}
\operatorname{Tr}
\left[
\bar c\,
\Delta_{\rm FP}(u)\,
c
\right].
\end{equation}

For \(N=5\), the interacting color algebra is

\begin{equation}
\mathfrak{su}(5),
\end{equation}

which has dimension \(24\). Thus,

\begin{equation}
\Delta_{\rm FP}(u)
\end{equation}

acts on the \(24\)-dimensional adjoint color space.

Grassmann Gaussian integration gives

\begin{equation}
\int d\bar c\,dc\,
e^{iS_{\rm gh}^{(2)}}
\propto
\det\Delta_{\rm FP}(u).
\end{equation}

Along the physical modulus analyzed here, \(\Delta_{\rm FP}(u)\) is nondegenerate, so its determinant can be compared directly between the backgrounds.

On this local branch, the relative variation of

\begin{equation}
\det\Delta_{\rm FP}(u)
\end{equation}

remains at the level of numerical precision. The variation of the determinant-defined one-loop profile discussed in Section 4 therefore comes mainly from the bosonic hard sector and the fermionic sector.

\subsection{Majorana--Weyl fermions and the Pfaffian}\label{majoranaweyl-fermions-and-the-pfaffian}

The fermionic fluctuations couple to \(A_\mu(u)\) through the background covariant derivative \(D_\mu\). In this subsection, we specify the Majorana--Weyl gamma-matrix representation, basis ordering, and Pfaffian orientation used in the one-loop calculation.

\subsubsection{Majorana--Weyl gamma-matrix convention}\label{majoranaweyl-gamma-matrix-convention}

We use the real \(16\)-dimensional positive-chirality Majorana--Weyl carrier

\begin{equation}
S_+\simeq\mathbb R^{16}
\end{equation}

with a fixed real basis.

We denote the one-index chiral blocks used in the calculation by

\begin{equation}
\widehat\gamma^\mu
\in
M_{16}(\mathbb R),
\qquad
\mu=0,\ldots,9.
\end{equation}

They are obtained from a \(32\)-component Clifford representation as

\begin{equation}
\widehat\gamma^\mu
=
-\frac12
V_+^{\mathsf T}
C\Gamma_{32}^{\mu}
V_+,
\qquad
\Gamma_{\rm chir}V_+=V_+.
\end{equation}

Thus, the charge-conjugation matrix \(C\) has already been incorporated into the chiral blocks

\begin{equation}
\widehat\gamma^\mu_{\alpha\beta},
\end{equation}

with both spinor indices lowered. No additional \(16\times16\) charge-conjugation matrix is applied when constructing the fermion matrix.

All these blocks are real and symmetric,

\begin{equation}
(\widehat\gamma^\mu)^{\mathsf T}
=
\widehat\gamma^\mu,
\qquad
\widehat\gamma^0=I_{16}.
\end{equation}

We use the same mostly-plus Lorentz metric as in the main text,

\begin{equation}
\eta^{\mu\nu}
=
\operatorname{diag}(-1,+1,\ldots,+1).
\end{equation}

Because \(\widehat\gamma^\mu\) are same-chirality, charge-conjugated one-index blocks, the Clifford relation is not written as a simple anticommutator of two identical types of blocks.

With no summation over \(\mu\), we define

\begin{equation}
\widehat{\widetilde\gamma}^{\,\mu}
:=
\eta_{\mu\mu}\widehat\gamma^\mu.
\end{equation}

The exact mixed chiral Clifford relations are then

\begin{equation}
\widehat\gamma^\mu
\widehat{\widetilde\gamma}^{\,\nu}
+
\widehat\gamma^\nu
\widehat{\widetilde\gamma}^{\,\mu}
=
2\eta^{\mu\nu}I_{16},
\end{equation}

and

\begin{equation}
\widehat{\widetilde\gamma}^{\,\mu}
\widehat\gamma^\nu
+
\widehat{\widetilde\gamma}^{\,\nu}
\widehat\gamma^\mu
=
2\eta^{\mu\nu}I_{16}.
\end{equation}

In particular,

\begin{equation}
\widehat{\widetilde\gamma}^{\,0}
=
-I_{16},
\qquad
\widehat{\widetilde\gamma}^{\,i}
=
\widehat\gamma^i,
\qquad
i=1,\ldots,9.
\end{equation}

The relation to the notation used in the main text is

\begin{equation}
\widehat\gamma^\mu:=i\gamma^\mu.
\end{equation}

With this notation, the main-text fermionic operator

\begin{equation}
\mathcal M_F
=
i\gamma^\mu D_\mu
\end{equation}

is represented explicitly as

\begin{equation}
\mathcal M_F
=
\sum_{\mu=0}^{9}
\widehat\gamma^\mu\otimes D_\mu.
\end{equation}

Thus, the notation used in the main text and the explicit real matrix representation in this Appendix define the same fermionic quadratic operator.

\subsubsection{Explicit realization}\label{explicit-realization}

For reproducibility, we record the \(16\)-component row and column ordering and the explicit matrix representation used in the one-loop calculation.

The spinor basis is ordered lexicographically as

\begin{equation}
|0000\rangle,
|0001\rangle,
\ldots,
|1111\rangle,
\end{equation}

with the first bit varying most slowly.

Define

\begin{equation}
I=
\begin{pmatrix}
1&0\\
0&1
\end{pmatrix},
\qquad
X=
\begin{pmatrix}
0&1\\
1&0
\end{pmatrix},
\end{equation}

\begin{equation}
Z=
\begin{pmatrix}
1&0\\
0&-1
\end{pmatrix},
\qquad
J=
\begin{pmatrix}
0&1\\
-1&0
\end{pmatrix}.
\end{equation}

In this ordering, the gamma blocks are

\begin{equation}
\begin{aligned}
\widehat\gamma^0
&=
I\otimes I\otimes I\otimes I,
\\
\widehat\gamma^1
&=
-I\otimes Z\otimes I\otimes Z,
\\
\widehat\gamma^2
&=
Z\otimes Z\otimes J\otimes J,
\\
\widehat\gamma^3
&=
J\otimes J\otimes I\otimes Z,
\\
\widehat\gamma^4
&=
X\otimes Z\otimes J\otimes J,
\\
\widehat\gamma^5
&=
I\otimes X\otimes X\otimes Z,
\\
\widehat\gamma^6
&=
J\otimes X\otimes J\otimes Z,
\\
\widehat\gamma^7
&=
-I\otimes X\otimes Z\otimes Z,
\\
\widehat\gamma^8
&=
J\otimes I\otimes I\otimes J,
\\
\widehat\gamma^9
&=
-I\otimes I\otimes I\otimes X.
\end{aligned}
\end{equation}

Every matrix entry is exactly \(0\) or \(\pm1\). Together with the basis ordering above, these tensor-product formulas uniquely specify the real \(16\times16\) matrices used in the one-loop calculation.

In color space, we normalize the traceless Hermitian generators by

\begin{equation}
\operatorname{Tr}(T_aT_b)
=
\delta_{ab}
\end{equation}

and use the fixed ordering

\begin{equation}
\begin{aligned}
(&S_{12},A_{12},S_{13},A_{13},S_{14},A_{14},S_{15},A_{15},
\\
& S_{23},A_{23},S_{24},A_{24},S_{25},A_{25},
\\
& S_{34},A_{34},S_{35},A_{35},S_{45},A_{45},
H_1,H_2,H_3,H_4).
\end{aligned}
\end{equation}

Here

\begin{equation}
S_{ij}
=
\frac{E_{ij}+E_{ji}}{\sqrt2},
\qquad
A_{ij}
=
\frac{-iE_{ij}+iE_{ji}}{\sqrt2},
\end{equation}

and

\begin{equation}
H_k
=
\frac{
\operatorname{diag}
(\underbrace{1,\ldots,1}_{k},-k,0,\ldots,0)
}{
\sqrt{k(k+1)}
},
\qquad
k=1,\ldots,4.
\end{equation}

In this basis,

\begin{equation}
D_\mu X
=
i[A_\mu(u),X]
\end{equation}

is represented by a real \(24\times24\) matrix satisfying

\begin{equation}
D_\mu^{\mathsf T}
=
-D_\mu.
\end{equation}

\subsubsection{Fermion matrix and Pfaffian orientation}\label{fermion-matrix-and-pfaffian-orientation}

The fermionic fluctuation space has dimension

\begin{equation}
16\times24=384.
\end{equation}

The tensor-product ordering used in the calculation takes the spinor index as the outer, slowly varying index and the adjoint index as the inner, rapidly varying index,

\begin{equation}
\mathcal B_F
=
(
e_1\otimes T_1,\ldots,e_1\otimes T_{24},
e_2\otimes T_1,\ldots,e_{16}\otimes T_{24}
).
\end{equation}

The flattened index is

\begin{equation}
I
=
24(\alpha-1)+(a-1),
\qquad
I=0,\ldots,383,
\quad
\alpha=1,\ldots,16,
\quad
a=1,\ldots,24.
\end{equation}

In this ordered real basis, the fermionic quadratic operator is

\begin{equation}
\mathcal M_F(u)
=
\sum_{\mu=0}^{9}
\widehat\gamma^\mu\otimes D_\mu(u)
=
i\gamma^\mu D_\mu.
\end{equation}

Using

\begin{equation}
(\widehat\gamma^\mu)^{\mathsf T}
=
\widehat\gamma^\mu
\end{equation}

and

\begin{equation}
D_\mu^{\mathsf T}
=
-D_\mu,
\end{equation}

we directly obtain

\begin{equation}
\mathcal M_F(u)^{\mathsf T}
=
-\mathcal M_F(u),
\qquad
\mathcal M_F(u)
\in
\mathfrak{so}(384,\mathbb R).
\end{equation}

The quadratic fermionic action is

\begin{equation}
S_F^{(2)}
=
\frac{1}{2g^2}
\psi^{\mathsf T}
\mathcal M_F(u)
\psi.
\end{equation}

The Grassmann Gaussian integral gives

\begin{equation}
\int d\psi\,
e^{iS_F^{(2)}}
\propto
\operatorname{Pf}\mathcal M_F(u),
\end{equation}

with

\begin{equation}
\bigl[
\operatorname{Pf}\mathcal M_F(u)
\bigr]^2
=
\det\mathcal M_F(u).
\end{equation}

The Pfaffian orientation is fixed by the exterior orientation associated with the ordered basis \(\mathcal B_F\),

\begin{equation}
d\psi_{1,1}
\wedge\cdots\wedge
d\psi_{1,24}
\wedge
d\psi_{2,1}
\wedge\cdots\wedge
d\psi_{16,24}.
\end{equation}

The same basis ordering and orientation are used for all \(23\) analyzed backgrounds. Numerically, throughout these backgrounds we find

\begin{equation}
\operatorname{rank}\mathcal M_F
=
384,
\qquad
\operatorname{sign}
\operatorname{Pf}\mathcal M_F
=
+1,
\qquad
\arg
\operatorname{Pf}\mathcal M_F
=
0.
\end{equation}

The \(u\)-dependence of

\begin{equation}
\operatorname{Pf}\mathcal M_F(u)
\end{equation}

is therefore compared using the same gamma-matrix representation, basis ordering, and Pfaffian orientation at every point. It measures the change of the fermionic quadratic response to the microscopic matrix structure of the background.

\subsubsection{Gauge and proper-Lorentz covariance of the Pfaffian magnitude}\label{gauge-and-lorentz-covariance-of-pfaffian}

The Pfaffian magnitude used in the main text has a simpler invariant status than the auxiliary hard-boson determinant. Let \(U\) be a color-gauge transformation on the interacting \(SU(5)\) adjoint space and let \(\Lambda\in SO^+(9,1)\) be a connected proper-Lorentz transformation with Majorana--Weyl spin representative \(S(\Lambda)\). The gamma-matrix intertwining relation implies that the fermion bilinear is related by a congruence transformation,

\begin{equation}
\mathcal M_F(\Lambda A^U)
=
T^{-\mathsf T}\,
\mathcal M_F(A)
\,T^{-1},
\qquad
T=S(\Lambda)\otimes R_{\rm adj}(U),
\end{equation}

up to the equivalent inverse convention obtained by transforming the fermionic coefficient vector in the opposite direction. Here \(R_{\rm adj}(U)\) is the real adjoint representation on the \(24\)-dimensional interacting color space.

For connected transformations, both representation determinants are unity,

\begin{equation}
\det S(\Lambda)=1,
\qquad
\det R_{\rm adj}(U)=1,
\end{equation}

because the corresponding Lie-algebra representation matrices are traceless and the identity component is connected. Hence

\begin{equation}
\det T=1.
\end{equation}

Using \(\operatorname{Pf}(B^{\mathsf T}MB)=\det(B)\operatorname{Pf}(M)\), we obtain

\begin{equation}
|\operatorname{Pf}\mathcal M_F(\Lambda A^U)|
=
|\operatorname{Pf}\mathcal M_F(A)|.
\end{equation}

Thus \(\mathcal P_F=\log|\operatorname{Pf}\mathcal M_F|\) is a scalar on the local quotient by color gauge transformations and connected proper-Lorentz transformations. The absolute value avoids any dependence on an orientation convention under transformations outside the connected component; the present physical quotient uses the connected proper-Lorentz frame directions.

\subsection{Fixed-section determinant diagnostic}\label{determinantpfaffian-contribution}

Combining the three quadratic sectors, the local Gaussian factor, up to a normalization independent of \(u\), is

\begin{equation}
Z_{\rm det}^{(1)}(u)
\propto
e^{i\phi_B}
\frac{
\det\Delta_{\rm FP}(u)\,
\operatorname{Pf}\mathcal M_F(u)
}{
\sqrt{
|\det H_{\rm hard}(u)|
}
}.
\end{equation}

Here \(\phi_B\) is the phase associated with the Lorentzian bosonic Gaussian. Its derivation is given in Appendix D.

To describe the combined Gaussian variation in the chosen Lorentz section, we define the fixed-section determinant diagnostic by

\begin{equation}
\Gamma_{1,\mathrm{sec}}^{\det}(u)
=
\frac12
\log
|\det H_{\rm hard}(u)|
-
\log
|\det\Delta_{\rm FP}(u)|
-
\log
|\operatorname{Pf}\mathcal M_F(u)|.
\end{equation}

Relative to the reference background \(u=0\), we define

\begin{equation}
\Delta\Gamma_{1,\mathrm{sec}}^{\det}(u)
=
\Gamma_{1,\mathrm{sec}}^{\det}(u)
-
\Gamma_{1,\mathrm{sec}}^{\det}(0).
\end{equation}

The quantity \(\Gamma_{1,\mathrm{sec}}^{\det}(u)\) combines the quadratic one-loop response of the \(169\) hard bosonic directions, the \(24\)-dimensional Faddeev--Popov ghost sector, and the \(384\)-dimensional Majorana--Weyl fermionic sector in the chosen Lorentz section and auxiliary coefficient-space normalization. It is useful for decomposing the finite-dimensional Gaussian factors, but it is not an intrinsic scalar on the proper-Lorentz quotient.

On the analyzed local branch, this fixed-section diagnostic varies with \(u\). The intrinsic quotient-scalar statement is instead the nonconstant Pfaffian magnitude established above. Combining the fixed-section Gaussian factors with the collective-coordinate quotient Jacobian gives the representative-independent local one-loop density magnitude and \(W_{\rm rel}(u)\) used in Section 4. This construction is given in Appendix D.

\section{Lorentzian Gaussian and local one-loop density}\label{lorentzian-gaussian-and-local-one-loop-density}

\subsection{Definition of the Lorentzian Gaussian}\label{definition-of-the-lorentzian-gaussian}

In Appendix C, we constructed the quadratic operators for the bosonic fluctuations transverse to the physical modulus, the Faddeev--Popov ghosts, and the Majorana--Weyl fermions. In this Appendix, we combine their Gaussian factors with the measure along the physical modulus and construct the local one-loop density used in Section 4.

Let the quadratic form in the bosonic hard sector be

\begin{equation}
\frac12
x^{\mathsf T}
H_{\rm hard}(u)
x,
\end{equation}

where, at \(N=5\),

\begin{equation}
x\in\mathbb R^{169}.
\end{equation}

We define the finite-dimensional Lorentzian Gaussian by adding an infinitesimal positive damping term in this real fluctuation coefficient space,

\begin{equation}
\lim_{\epsilon\to0^+}
\int_{\mathbb R^{169}}
d^{169}x\,
\exp
\left[
\frac{i}{2}
x^{\mathsf T}
H_{\rm hard}(u)
x
-
\frac{\epsilon}{2}
x^{\mathsf T}x
\right].
\end{equation}

The positive-definite quadratic form

\begin{equation}
x^{\mathsf T}x
\end{equation}

appearing in the damping term acts only on the finite-dimensional real coefficient space of the hard fluctuations. It is an auxiliary prescription used to define the boundary value and phase of the oscillatory Gaussian. It does not replace the Lorentzian contractions in the action and is not a Wick rotation of the background.

For one positive eigenvalue of \(H_{\rm hard}\), the Gaussian contributes the phase

\begin{equation}
e^{i\pi/4},
\end{equation}

while one negative eigenvalue contributes

\begin{equation}
e^{-i\pi/4}.
\end{equation}

On the analyzed branch, the hard-sector inertia is

\begin{equation}
(95,74).
\end{equation}

The common bosonic Gaussian phase is therefore

\begin{equation}
e^{i\phi_B}
=
\exp
\left[
\frac{i\pi}{4}(95-74)
\right]
=
e^{-3\pi i/4}.
\end{equation}

Since the numbers of positive and negative hard directions remain unchanged along the physical modulus, this phase is common to the analyzed branch. The relative magnitude of the bosonic hard contribution is therefore determined by

\begin{equation}
|\det H_{\rm hard}(u)|^{-1/2}.
\end{equation}

The construction above defines a local fluctuation integral around each real background. The extension in the homogeneous radial direction and its distinction from the global Lorentzian matrix-space integration problem are discussed in Appendix F.

\subsection{Measure along the physical modulus}\label{measure-along-the-physical-modulus}

The physical zero mode found in Appendix B is tangent to the classical solution curve

\begin{equation}
u\longmapsto A_\mu(u).
\end{equation}

We separate this direction from the Gaussian fluctuations and retain \(u\) as a collective coordinate.

To define tangent-vector normalization and local Jacobians, we use the auxiliary positive-definite inner product on the real coefficient space of the bosonic matrix components,

\begin{equation}
\langle a,b\rangle_{\rm coeff}
=
\delta^{\mu\nu}
\operatorname{Tr}(a_\mu b_\nu).
\end{equation}

Here \(\delta^{\mu\nu}\) is a positive-definite metric on the finite-dimensional coefficient space. It is used only to define lengths, orthogonal projections, and local Jacobians in that space. It is distinct from the Lorentzian metric \(\eta^{\mu\nu}\), which enters the action, the constraint, and the physical Lorentz contractions.

The corresponding line element along the solution curve is

\begin{equation}
d\ell
=
\sqrt{G_{uu}(u)}\,du,
\end{equation}

where

\begin{equation}
G_{uu}(u)
=
\delta^{\mu\nu}
\operatorname{Tr}
\left(
\frac{dA_\mu}{du}
\frac{dA_\nu}{du}
\right).
\end{equation}

Thus,

\begin{equation}
\sqrt{G_{uu}(u)}
\end{equation}

is the local measure factor that converts the normalized zero-mode coordinate into the branch coordinate \(u\).

The fixed-\(Q_\rho\) condition is

\begin{equation}
Q_\rho[A]
=
\operatorname{Tr}(A_\mu A^\mu)
-
Q_{\rm ref}.
\end{equation}

Its variation is

\begin{equation}
\delta Q_\rho
=
2\operatorname{Tr}
\left(
A^\mu\delta A_\mu
\right).
\end{equation}

If the configuration-space integral is restricted by inserting

\begin{equation}
\delta(Q_\rho[A]-9),
\end{equation}

the induced measure on the \(Q_\rho=9\) surface contains the factor

\begin{equation}
\|\nabla_{\rm coeff}Q_\rho\|_{\rm coeff}^{-1}.
\end{equation}

Measured with the auxiliary coefficient-space metric,

\begin{equation}
\|\nabla_{\rm coeff}Q_\rho\|_{\rm coeff}
=
2
\left[
\delta^{\mu\nu}
\operatorname{Tr}(A_\mu A_\nu)
\right]^{1/2}.
\end{equation}

This is a positive coefficient-space magnitude associated with the restriction to the constraint surface. It should not be confused with the Lorentzian contraction

\begin{equation}
\operatorname{Tr}(A_\mu A^\mu)
\end{equation}

that defines \(Q_\rho\).

To make the measure bookkeeping explicit, consider first the local finite-dimensional bosonic measure

\begin{equation}
d^{240}A\,\delta\!\left(Q_\rho[A]-9\right).
\end{equation}

At a regular point of the analyzed branch, choose local coefficient-space coordinates adapted to the color-gauge orbit, the Lorentz-frame orbit, the physical modulus, and the hard transverse directions,

\begin{equation}
(\theta_G,\theta_L,u,x)
\in
\mathbb R^{24}\times\mathbb R^{45}\times\mathbb R\times\mathbb R^{169}.
\end{equation}

Up to an overall normalization independent of the background, the local change of variables gives

\begin{equation}
d^{240}A\,\delta\!\left(Q_\rho[A]-9\right)
=
\sqrt{G_{uu}}
\frac{J_GJ_L}{J_Q}
\,d^{24}\theta_G\,d^{45}\theta_L\,du\,d^{169}x.
\end{equation}

This is a local finite-dimensional change-of-variables formula in the auxiliary coefficient-space metric. It does not define a finite global Lorentz-group volume, and it does not modify the Lorentzian contractions in the action. A nonperturbative Faddeev--Popov treatment of the divergent noncompact Lorentz volume in the full Lorentzian IIB matrix integral has been proposed in Ref.~\cite{Asano:2025}. The present construction is different: it is a local quotient of tangent directions near the constrained branch, used to define physical deformations and the associated local density. After removing the gauge- and frame-orbit coordinates, the geometric quotient measure on the modulus and hard directions is therefore

\begin{equation}
J_{\rm mod}^{\rm geom}(u)\,du\,d^{169}x,
\qquad
J_{\rm mod}^{\rm geom}(u)
:=
\sqrt{G_{uu}(u)}\frac{J_G(u)J_L(u)}{J_Q(u)}.
\end{equation}

We next separate the color-gauge and Lorentz-frame directions. Let \(V(u)\) be the matrix whose columns are the \(24\) independent color-gauge tangent vectors. Let \(R(u)\) contain the \(45\) Lorentz-frame tangent vectors after removing the gauge directions, and let

\begin{equation}
P_G(u)
\end{equation}

denote the projector onto the gauge-horizontal subspace.

The corresponding local factors are

\begin{equation}
J_G(u)
=
\sqrt{
\det
\left[
V(u)^{\mathsf T}V(u)
\right]
},
\end{equation}

\begin{equation}
J_L(u)
=
\sqrt{
\det
\left[
R(u)^{\mathsf T}
P_G(u)
R(u)
\right]
},
\end{equation}

and

\begin{equation}
J_Q(u)
=
\|\nabla_{\rm coeff}Q_\rho\|_{\rm coeff}.
\end{equation}

There are two equivalent ways to represent the color-gauge part of this local quotient measure. In a ghost-absorbed geometric convention, one uses \(J_{\rm mod}^{\rm geom}\) directly and does not multiply by another raw Faddeev--Popov determinant. In the explicit-FP convention used in Section 4, the background-gauge condition is accompanied by \(\det\Delta_{\rm FP}\), and the collective factor is defined with the compensating inverse determinant. The collective-coordinate factor is therefore

\begin{equation}
J_{\rm mod}(u)
=
\sqrt{G_{uu}(u)}
\,
\frac{
\sqrt{\det[V(u)^{\mathsf T}V(u)]}\,
\sqrt{\det[R(u)^{\mathsf T}P_G(u)R(u)]}
}{
\|\nabla_{\rm coeff}Q_\rho\|_{\rm coeff}\,
\det\Delta_{\rm FP}(u)
}.
\end{equation}

Equivalently,

\begin{equation}
J_{\rm mod}(u)
=
\sqrt{G_{uu}(u)}
\frac{
J_G(u)J_L(u)
}{
J_Q(u)\det\Delta_{\rm FP}(u)
}.
\end{equation}

It follows that

\begin{equation}
J_{\rm mod}(u)\det\Delta_{\rm FP}(u)
=
\sqrt{G_{uu}(u)}
\frac{
J_G(u)J_L(u)
}{
J_Q(u)
}.
\end{equation}

Thus, the two conventions give the same local quotient measure, and the gauge-orbit factor is counted only once. The \(169\)-dimensional hard Hessian is evaluated only after the gauge, constraint-normal, Lorentz-frame, and modulus directions have been removed. The explicit ghost determinant is therefore a representation of the gauge-orbit Jacobian in the chosen bookkeeping convention, not a second removal of the hard directions.

On the accepted branch analyzed here, \(\Delta_{\rm FP}(u)\) is numerically full rank, and

\begin{equation}
\det\Delta_{\rm FP}(u)>0
\end{equation}

throughout the sampled backgrounds. Therefore, within this explicit-FP convention, \(J_{\rm mod}(u)\) can be treated as a positive local measure factor on the analyzed branch, with no additional FP sign ambiguity in the density magnitude reported below.

\subsection{Local one-loop density magnitude}\label{local-one-loop-density-magnitude}

The fixed-section determinant diagnostic defined in Section 4 is

\begin{equation}
\Gamma_{1,\mathrm{sec}}^{\det}(u)
=
\frac12
\log|\det H_{\rm hard}(u)|
-
\log|\det\Delta_{\rm FP}(u)|
-
\log|\operatorname{Pf}\mathcal M_F(u)|.
\end{equation}

On the fixed-\(Q_\rho\) slice

\begin{equation}
Q_\rho=9,
\end{equation}

we define the local one-loop density magnitude including the collective-coordinate measure by

\begin{equation}
\mathcal W_{\rm 1loop}(u,9)
=
J_{\rm mod}(u)
\exp[-\Gamma_{1,\mathrm{sec}}^{\det}(u)].
\end{equation}

Equivalently,

\begin{equation}
\mathcal W_{\rm 1loop}(u,9)
=
J_{\rm mod}(u)
\frac{
|\det\Delta_{\rm FP}(u)|\,
|\operatorname{Pf}\mathcal M_F(u)|
}{
\sqrt{|\det H_{\rm hard}(u)|}
}.
\end{equation}

With the action convention of Section 2,

\begin{equation}
S_B[A]
=
\frac{1}{4g^2}
\operatorname{Tr}
\left(
F_{\mu\nu}F^{\mu\nu}
\right),
\end{equation}

the local Lorentzian contribution along the physical modulus can be written, up to an overall background-independent normalization, as

\begin{equation}
\mathcal N\,
e^{iS_B[A(u)]}
e^{-3\pi i/4}
\mathcal W_{\rm 1loop}(u,9)\,du.
\end{equation}

The classical action and the bosonic Gaussian phase are common along the analyzed physical modulus. They therefore cancel in comparisons of the local magnitude between backgrounds.

We define the relative local one-loop density magnitude with respect to the reference background \(u=0\) by

\begin{equation}
W_{\rm rel}(u)
=
\frac{
\mathcal W_{\rm 1loop}(u,9)
}{
\mathcal W_{\rm 1loop}(0,9)
}.
\end{equation}

Using

\begin{equation}
\Delta\Gamma_{1,\mathrm{sec}}^{\det}(u)
=
\Gamma_{1,\mathrm{sec}}^{\det}(u)
-
\Gamma_{1,\mathrm{sec}}^{\det}(0),
\end{equation}

this becomes

\begin{equation}
W_{\rm rel}(u)
=
\frac{
J_{\rm mod}(u)
}{
J_{\rm mod}(0)
}
\exp
\left[
-\Delta\Gamma_{1,\mathrm{sec}}^{\det}(u)
\right].
\end{equation}

This expression combines the quadratic fluctuations transverse to the modulus with the local measure along the modulus.

The quantity that is invariant under a reparametrization of the physical modulus is the corresponding density one-form. If \(u\) and \(v\) are two local coordinates on the same solution curve and the density coefficients in the two coordinates are denoted by \(\mathcal W_u\) and \(\mathcal W_v\), then

\begin{equation}
\mathcal W_u(u)\,du
=
\mathcal W_v(v)\,dv.
\end{equation}

The numerical profile \(W_{\rm rel}(u)\) quoted in this paper is the coefficient ratio in the fixed continuation coordinate \(u\). Because a positive one-dimensional density coefficient can be flattened locally by a suitable reparametrization, the nonconstancy of \(W_{\rm rel}(u)\) by itself is not an intrinsic scalar statement about the branch. The intrinsic scalar one-loop discrimination used in the main argument is instead the variation of \(\mathcal P_F=\log|\operatorname{Pf}\mathcal M_F|\). The fixed-section determinant diagnostic \(\Gamma_{1,\mathrm{sec}}^{\det}\) is kept separate from this statement.

The representative independence of the complete quotient density can also be checked directly. We apply the same finite boost in the \((0,1)\) plane to every one of the \(23\) backgrounds and reconstruct the physical quotient, hard Hessian, Faddeev--Popov operator, fermion matrix, and geometric Jacobian from the transformed representatives. For rapidities \(0.1\), \(0.5\), and \(1.0\), the root-referenced Pfaffian profile and the complete density profile agree with the unboosted profiles to at most about \(3.2\times10^{-13}\). By contrast, the root-referenced fixed-section determinant diagnostic changes by up to approximately \(1.01\times10^{-2}\) at rapidity \(1.0\). The change is compensated by the geometric quotient Jacobian in the full density. This numerical covariance check is included in the reproducibility supplement.

Equivalently, the coefficient with respect to the auxiliary proper-length element

\begin{equation}
d\ell
=
\sqrt{G_{uu}}\,du
\end{equation}

may be written as

\begin{equation}
\mathcal W_\ell
=
\frac{
\mathcal W_u
}{
\sqrt{G_{uu}}
}.
\end{equation}

Normalizing this coefficient to the reference background gives

\begin{equation}
W_{\rm rel}^{(\ell)}(u)
=
W_{\rm rel}(u)
\sqrt{\frac{G_{uu}(0)}{G_{uu}(u)}}.
\end{equation}

The proper length here is defined by the auxiliary positive-definite coefficient-space inner product used for basis normalization and local projections. It is not a Lorentzian spacetime distance. Neither \(W_{\rm rel}(u)\) nor \(W_{\rm rel}^{(\ell)}(u)\) is a globally normalized probability. The invariant local object under reparametrization is the density one-form.

\subsection{Evaluation on the classical solution family}\label{evaluation-on-the-classical-solution-family}

We finally check that the leading one-loop profile can be evaluated on the classical constrained family \(A_\mu(u)\).

Factor a common positive background scale as

\begin{equation}
A_\mu=b_{\rm sc}\,\widehat A_\mu,
\qquad b_{\rm sc}>0,
\end{equation}

and define

\begin{equation}
\epsilon_g
=
\frac{g}{b_{\rm sc}^2}.
\end{equation}

This scale factor is used only for local loop counting; it is not the historical scale variable introduced in Appendix F. Let \(z\) denote local physical coordinates transverse to the modulus. We write the classical action locally as

\begin{equation}
S_B(u,z)
=
\epsilon_g^{-2}
\widehat S_0(u,z),
\end{equation}

with

\begin{equation}
\widehat S_0(u,z)
=
\widehat S_{0,*}
+
\frac12
z^{\mathsf T}
H(u)
z
+\cdots.
\end{equation}

Here \(H(u)\) is the transverse Hessian in the coefficient normalization used in the one-loop calculation.

The following local power counting assumes a neighborhood in which the hard block remains invertible, the Faddeev--Popov operator remains nonsingular with fixed sign, the Majorana--Weyl Pfaffian remains nonzero on a continuously chosen branch, and the corresponding finite-dimensional operators depend smoothly on the local coordinates. The pointwise gaps, ranks, and Pfaffian-sign checks on the accepted backgrounds are consistent with these assumptions on the analyzed local branch, but they are not an interval-arithmetic proof on an open neighborhood.

Let \(\Phi_1(u,z)\) denote the smooth \(O(1)\) logarithmic one-loop term in a chosen system of local quotient coordinates. Depending on the bookkeeping convention, \(\Phi_1\) includes the determinant factors together with the corresponding local Jacobian terms. The following argument uses only its smoothness and order in \(\epsilon_g\); it does not assign invariant meaning to the location of a stationary point of a density coefficient under an arbitrary change of local coordinates.

In such a fixed local coordinate system, the transverse stationary condition gives

\begin{equation}
z_*(u)
=
-
\epsilon_g^2
H(u)^{-1}
\partial_z
\Phi_1(u,0)
+
O(\epsilon_g^4).
\end{equation}

Hence, whenever the corresponding local corrected transverse stationary point exists smoothly in that coordinate description, its displacement starts at order

\begin{equation}
O(\epsilon_g^2).
\end{equation}

The leading one-loop quantities can therefore be evaluated on the classical constrained family \(z=0\) at this order. This is only a local loop-counting statement. It does not prove the existence or uniqueness of a full quantum-corrected saddle, does not define a quotient-invariant corrected position, and does not select a position along the modulus \(u\).

The notation \(\widehat S_0\) is used only for this local loop-counting argument and does not change the action convention of Section 2.

\subsection{Numerical one-loop profile}\label{numerical-one-loop-profile}

On the analyzed local branch, both the quadratic fluctuation factors and the collective-coordinate measure depend on \(u\). Their combination,

\begin{equation}
W_{\rm rel}(u),
\end{equation}

is therefore also nonconstant.

For the fixed-section combined determinant diagnostic alone, the sampled maximum-to-minimum factor is approximately

\begin{equation}
1.32.
\end{equation}

The intrinsic Pfaffian magnitude has the smaller sampled maximum-to-minimum ratio

\begin{equation}
1.0520.
\end{equation}

After including the collective-coordinate measure, the \(23\) sampled accepted backgrounds give

\begin{equation}
0.807
\lesssim
W_{\rm rel}(u)
\lesssim
1.032.
\end{equation}

These extrema, as well as the maximum-to-minimum ratio below, are sampled values on the \(23\) accepted backgrounds. They are not certified extrema over a continuous interval or over a global branch.

The corresponding sampled ratio is

\begin{equation}
\frac{
\max W_{\rm rel}
}{
\min W_{\rm rel}
}
\simeq
1.28.
\end{equation}

The logarithmic branch spans relative to the reference background \(u=0\) are approximately

\begin{equation}
0.225
\qquad
\text{for the hard bosonic sector},
\end{equation}

\begin{equation}
0.0507
\qquad
\text{for the Majorana--Weyl fermionic sector},
\end{equation}

and

\begin{equation}
0.0297
\qquad
\text{for the collective-coordinate measure}.
\end{equation}

The relative variation of the Faddeev--Popov determinant remains at the level of

\begin{equation}
10^{-14}.
\end{equation}

Thus, on this branch, the largest variation comes from the hard bosonic fluctuations. The Majorana--Weyl Pfaffian and the collective-coordinate measure also vary along the same physical modulus, while the Faddeev--Popov determinant remains numerically constant to the accuracy quoted above.

Combining these contributions, the fixed-\(u\) coefficient \(W_{\rm rel}(u)\) is nonconstant along the analyzed physical modulus. Using the auxiliary proper-length coefficient defined above, the same \(23\) backgrounds give

\begin{equation}
0.8068
\lesssim
W_{\rm rel}^{(\ell)}(u)
\lesssim
1.0320,
\end{equation}

and

\begin{equation}
\frac{\max W_{\rm rel}^{(\ell)}}{\min W_{\rm rel}^{(\ell)}}
\simeq
1.2790.
\end{equation}

Thus the density coefficient remains nonconstant in the auxiliary proper-length parametrization used in the calculation. This does not make density-coefficient nonconstancy a parametrization-independent statement; the invariant geometric object is the quotient density one-form. The intrinsic scalar discrimination of the physical quotient points is provided by the nonconstant Majorana--Weyl Pfaffian magnitude. The combined hard determinant remains a fixed-section diagnostic.

In Appendix F, we extend the fixed-\(Q_\rho\) density

\begin{equation}
\mathcal W_{\rm 1loop}(u,9)
\end{equation}

to nearby radial slices

\begin{equation}
s=Q_\rho
\end{equation}

and derive the exact factorization

\begin{equation}
\mathcal W_{\rm 1loop}(u,s)
=
b(s)^{92}
\mathcal W_{\rm 1loop}(u,9).
\end{equation}

\section{Framed representation lift and one-loop comparison}\label{framed-representation-lift-and-one-loop-comparison}

\subsection{SO(5) structural frame and color-gauge covariance}\label{so5-structural-frame-and-color-gauge-covariance}

In Appendix A, we constructed the general-\(L\) representation probe from the

\begin{equation}
\mathbf{10}\oplus\mathbf{14}
\end{equation}

decomposition of the \(N=5\) background,

\begin{equation}
\widetilde A_\mu^{(L)}
=
\sum_{a<b}
g_\mu^{ab}T_{ab}^{(L)}
+
\mathcal E_L(S_\mu).
\end{equation}

This construction is defined relative to an embedding

\begin{equation}
\iota:
\mathfrak{so}(5)
\hookrightarrow
\mathfrak{su}(5),
\end{equation}

which specifies the SO(5) structural frame in color space.

The relevant starting point for the representation lift is therefore not the matrix configuration \(A_\mu\) alone, but the framed pair

\begin{equation}
(A,\iota).
\end{equation}

The embedding \(\iota\) specifies additional structural data used to resolve the matrix configuration into the \(\mathbf{10}\) and \(\mathbf{14}\) sectors. It is not an additional gauge symmetry of the IIB matrix model.

Under a color-gauge transformation,

\begin{equation}
A_\mu
\longmapsto
UA_\mu U^\dagger,
\qquad
U\in SU(5),
\end{equation}

we transport the structural frame simultaneously,

\begin{equation}
\iota
\longmapsto
\operatorname{Ad}_U\circ\iota.
\end{equation}

Relative to the original frame, write

\begin{equation}
A_\mu
=
\sum_{a<b}
g_\mu^{ab}T_{ab}^{(1)}
+
\mathcal E_1(S_\mu).
\end{equation}

Under the simultaneous transport,

\begin{equation}
T_{ab}^{(1)}
\longmapsto
UT_{ab}^{(1)}U^\dagger,
\end{equation}

and

\begin{equation}
\mathcal E_1(S)
\longmapsto
U\mathcal E_1(S)U^\dagger.
\end{equation}

Hence,

\begin{equation}
UA_\mu U^\dagger
=
\sum_{a<b}
g_\mu^{ab}
\left(
UT_{ab}^{(1)}U^\dagger
\right)
+
U\mathcal E_1(S_\mu)U^\dagger.
\end{equation}

The abstract coefficient data

\begin{equation}
g_\mu^{ab},
\qquad
S_\mu
\end{equation}

are therefore unchanged when they are read in the transported frame.

The general-\(L\) construction is obtained by coupling these same coefficient data to \(T_{ab}^{(L)}\) and \(\mathcal E_L\). With the canonical identification of the representation space \(V_L\), this gives

\begin{equation}
\widetilde A_\mu^{(L)}
\left[
UAU^\dagger,
\operatorname{Ad}_U\circ\iota
\right]
=
\widetilde A_\mu^{(L)}[A,\iota].
\end{equation}

The sector tensors

\begin{equation}
\mathsf X_{\mu\nu},
\qquad
\mathsf Y_{\mu\nu}
\end{equation}

are likewise unchanged. Consequently, the lifted Lorentzian extent operator \(K^{(L)}\), its spectrum, and the derived quantity \(R_{3+6}(L,u)\) are independent of the choice of color-gauge representative when the structural frame is transported together with the background.

Thus, representative independence holds on the \textbf{framed color-gauge orbit}, rather than on the matrix configuration considered without the SO(5) structural frame. This is the covariance property used in Section 5.

\subsection{\texorpdfstring{Residual frame equivalence of the higher-\(L\) lift}{Residual frame equivalence of the higher-L lift}}\label{residual-frame-equivalence-of-the-higher-l-lift}

Transformations that preserve the chosen SO(5) embedding are characterized by its normalizer. For the embedding used here,

\begin{equation}
N_{SU(5)}
\bigl(
\iota(SO(5))
\bigr)
=
Z_5\,SO(5).
\end{equation}

The embedded SO(5) transformations and the center of \(SU(5)\) therefore give residual equivalences of the same structural framing.

Now consider a generic transformation

\begin{equation}
U\notin
N_{SU(5)}
\bigl(
\iota(SO(5))
\bigr).
\end{equation}

If the matrices are conjugated,

\begin{equation}
A_\mu\longmapsto UA_\mu U^\dagger,
\end{equation}

while the structural frame \(\iota\) is kept fixed, the transformed matrices are decomposed again relative to the original SO(5) frame. The resulting \(\mathbf{10}\) and \(\mathbf{14}\) coefficient data are then mixed.

At \(L=1\), the ordinary quadratic trace combines the two sectors with the original relative weight, and the effect of this change is numerically negligible in the spectrum. For \(L>1\), however, the two sectors enter with the representation-dependent relative weight

\begin{equation}
\rho_L
=
\frac{4L(L+3)+5}{42}.
\end{equation}

The lifted spectrum therefore depends on whether the SO(5) frame is transported consistently.

This distinction is also visible numerically. If the background is conjugated while the structural frame is incorrectly held fixed, the discrepancies in the lifted spectra are of order

\begin{equation}
3.3\times10^{-2}
\ \text{to}\
1.1\times10^{-1}
\qquad
(L=2),
\end{equation}

and

\begin{equation}
5.9\times10^{-2}
\ \text{to}\
1.4\times10^{-1}
\qquad
(L=3).
\end{equation}

By contrast, in \(69\) numerical tests in which the structural frame was transported together with the background, the maximum discrepancies were

\begin{equation}
1.6\times10^{-15}
\qquad
(L=1),
\end{equation}

\begin{equation}
2.2\times10^{-15}
\qquad
(L=2),
\end{equation}

and

\begin{equation}
3.7\times10^{-15}
\qquad
(L=3).
\end{equation}

For \(R_{3+6}\), the maximum discrepancy was

\begin{equation}
4.4\times10^{-15}.
\end{equation}

These checks agree with the exact simultaneous-transport covariance described above.

The higher-\(L\) construction should therefore be regarded as a quantity defined on a

\begin{equation}
\text{color-gauge orbit equipped with a transported SO(5) structural frame}.
\end{equation}

No additional assumption is made that the physical color-gauge orbit itself uniquely determines an SO(5) embedding. In particular, the structural frame is part of the definition of this representation probe.

\subsection{Pointwise representation probe along the physical modulus}\label{pointwise-representation-probe-along-the-physical-modulus}

We now apply the framed decomposition and representation lift to the physical modulus

\begin{equation}
u\longmapsto A_\mu(u)
\end{equation}

constructed in Section 3.

All \(23\) accepted backgrounds admit the required

\begin{equation}
\mathbf{10}\oplus\mathbf{14}
\end{equation}

decomposition relative to the transported SO(5) structural frame. The maximum decomposition residual is approximately

\begin{equation}
1.4\times10^{-16}.
\end{equation}

At \(L=1\), the construction reproduces the original \(N=5\) matrices. The maximum matrix-reconstruction difference is approximately

\begin{equation}
3.3\times10^{-16},
\end{equation}

and the maximum spectral difference is approximately

\begin{equation}
3.0\times10^{-15}.
\end{equation}

Thus, at every point on the sampled branch, the same framed microscopic data

\begin{equation}
\left\{
g_\mu^{ab}(u),
S_\mu(u);
\iota
\right\}
\end{equation}

can be used as the common input for the \(L=1,2,3\) structural representation probes.

As shown in Appendix A, the \(L=1\) lifted \(K\)-spectrum remains common within numerical precision over the analyzed local interval. At \(L=2\) and \(L=3\), the same classical microscopic deformation becomes visible as a nonzero spectral response.

For the \(23\) sampled accepted backgrounds, the numerical variations are

\begin{longtable}[]{@{}
  >{\raggedright\arraybackslash}p{(\columnwidth - 4\tabcolsep) * \real{0.2727}}
  >{\raggedleft\arraybackslash}p{(\columnwidth - 4\tabcolsep) * \real{0.3636}}
  >{\raggedleft\arraybackslash}p{(\columnwidth - 4\tabcolsep) * \real{0.3636}}@{}}
\toprule\noalign{}
\begin{minipage}[b]{\linewidth}\raggedright
quantity
\end{minipage} & \begin{minipage}[b]{\linewidth}\raggedleft
\(L=2\)
\end{minipage} & \begin{minipage}[b]{\linewidth}\raggedleft
\(L=3\)
\end{minipage} \\
\midrule\noalign{}
\endhead
\bottomrule\noalign{}
\endlastfoot
maximum change of lifted spectrum & \(5.07\times10^{-4}\) & \(1.04\times10^{-3}\) \\
\(\operatorname{span}_u I_L\) & \(1.83\times10^{-4}\) & \(5.57\times10^{-4}\) \\
\(\operatorname{span}_u R_{3+6}(L,u)\) & \(7.30\times10^{-4}\) & \(1.82\times10^{-3}\) \\
\end{longtable}

The branchwise results themselves are numerical sampled-point statements.

For pointwise comparisons with the reference background \(u=0\), we define

\begin{equation}
\Delta I_L(u)
=
I_L(u)-I_L(0).
\end{equation}

For the variation over the sampled branch, we use

\begin{equation}
\operatorname{span}_u I_L
=
\max_u I_L-\min_u I_L.
\end{equation}

The first quantity is therefore a pointwise difference from the reference background, while the second summarizes the sampled variation over the branch.

The higher-\(L\) family does not change the underlying dynamical background. The dynamical matrix size remains

\begin{equation}
N=5.
\end{equation}

It only changes the representation-dependent relative weight with which the same SO(5)-resolved coefficient data are read. It is therefore a structural representation probe, not a family of higher-\(N\) classical solutions.

\subsection{Comparison with the intrinsic fermionic one-loop response}\label{comparison-with-the-one-loop-response}

The intrinsic one-loop scalar used for the comparison is

\begin{equation}
\Delta\mathcal P_F(u)
=
\Delta\log|\operatorname{Pf}\mathcal M_F(u)|.
\end{equation}

For the framed representation probe, define

\begin{equation}
\widehat I_L(u)
=
\frac{\operatorname{Tr}_{10}[(K^{(L)}(u))^2]}
{[\operatorname{Tr}_{10}K^{(L)}(u)]^2},
\qquad
\Delta\widehat I_L(u)=\widehat I_L(u)-\widehat I_L(0).
\end{equation}

On the same ordered set of \(23\) accepted backgrounds, the Pearson coefficients of \(\Delta\mathcal P_F\) with \(\Delta\widehat I_2\), \(\Delta\widehat I_3\), \(R_{3+6}(2,u)\), and \(R_{3+6}(3,u)\) are

\begin{equation}
0.99908,
\qquad
0.99943,
\qquad
0.99909,
\qquad
0.99922,
\end{equation}

respectively. These coefficients summarize co-variation along one short branch. They are not used as evidence for a causal relation, an exact functional identity, or a symmetry theorem, and they are not required for the main conclusion. In particular, no attempt is made to assign independent statistical significance to the Pearson values for these deterministically sampled points.

\begin{figure}[t]
\centering
\includegraphics[width=0.72\textwidth]{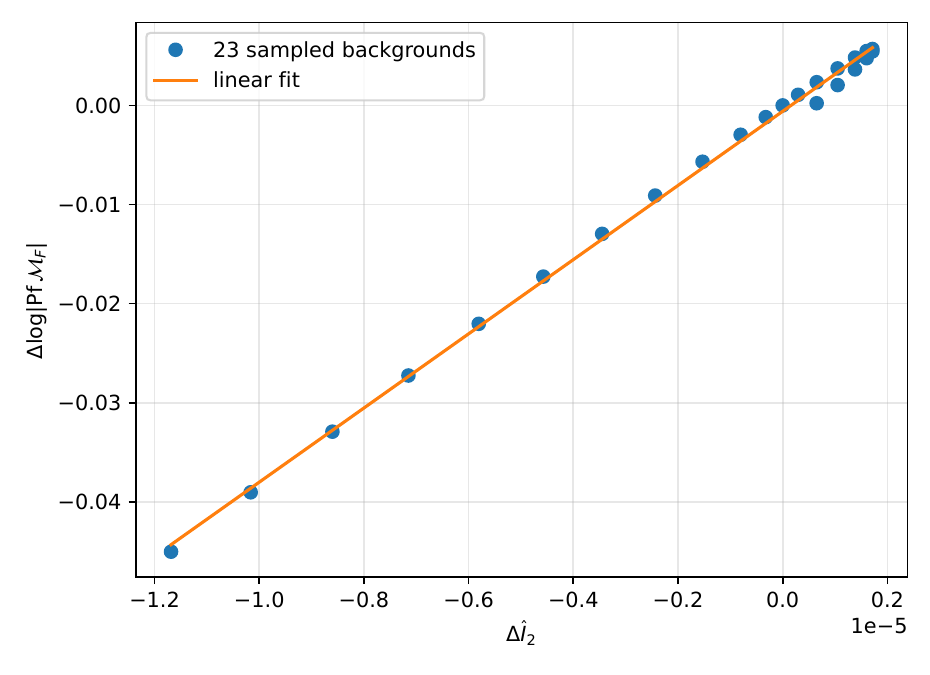}
\caption{Pointwise comparison between the intrinsic fermionic one-loop response $\Delta\mathcal P_F(u)=\Delta\log|\operatorname{Pf}\mathcal M_F|$ and the normalized framed response $\Delta\widehat I_2(u)$ on the same $23$ accepted backgrounds. The Pearson correlation is approximately $0.99908$; the line is a least-squares linear fit used only as a descriptive guide. The comparison is branch-local and does not imply a causal relation or an exact functional identity.}
\label{fig:correlation-one-loop-vs-I2}
\end{figure}

The fixed-chart density coefficient \(\log W_{\rm rel}\) and the fixed-section determinant diagnostic are retained in the reproducibility archive as separate quantities. They are not substituted for the intrinsic Pfaffian scalar in the comparison above.

\section{Radial continuation and stationary landscape}\label{radial-continuation-and-stationary-landscape}

In Sections 3--7, we analyzed a one-dimensional physical modulus on the normalization slice

\begin{equation}
Q_\rho=9.
\end{equation}

In this Appendix, we examine its homogeneous continuation to nearby radial slices, derive the exact radial scaling of the local one-loop density, and distinguish the resulting reduced radial problem from the unresolved global Lorentzian matrix integral. We also summarize the role of the historical algebraic special point and the broader stationary-solution landscape.

\subsection{Nearby radial continuation}\label{nearby-radial-continuation}

The quantity used to fix the homogeneous radial normalization is

\begin{equation}
Q_\rho[A]
=
\operatorname{Tr}(A_\mu A^\mu)
-
Q_{\rm ref},
\qquad
Q_{\rm ref}=6+2\sqrt3.
\end{equation}

On the slice used in Sections 3--6,

\begin{equation}
Q_\rho=9,
\end{equation}

so that

\begin{equation}
\operatorname{Tr}(A_\mu A^\mu)
=
9+Q_{\rm ref}
=
15+2\sqrt3
\equiv D_0.
\end{equation}

The contraction

\begin{equation}
\operatorname{Tr}(A_\mu A^\mu)
\end{equation}

is defined with the Lorentzian metric. It is not a positive-definite matrix norm. The choice \(Q_\rho=9\) fixes the homogeneous radial normalization of the analyzed branch; the value \(9\) itself is not assumed to be dynamically selected.

We now introduce

\begin{equation}
s:=Q_\rho
\end{equation}

and define a homogeneous continuation of the physical modulus by

\begin{equation}
A_\mu(u;s)
=
b(s)A_\mu(u;9).
\end{equation}

Under a homogeneous rescaling \(A_\mu\mapsto bA_\mu\),

\begin{equation}
Q_\rho[bA]
=
b^2
\left[
Q_\rho[A]+Q_{\rm ref}
\right]
-
Q_{\rm ref}.
\end{equation}

Requiring

\begin{equation}
Q_\rho[A(u;s)]=s
\end{equation}

therefore gives

\begin{equation}
b(s)^2
=
\frac{s+Q_{\rm ref}}
{9+Q_{\rm ref}},
\end{equation}

and hence

\begin{equation}
A_\mu(u;s)
=
\left(
\frac{s+Q_{\rm ref}}
{9+Q_{\rm ref}}
\right)^{1/2}
A_\mu(u;9).
\end{equation}

This scaling relation is exact wherever the positive-real scaling factor is defined. The physical certification used in this paper is more limited: it concerns a neighborhood of \(s=9\), not the full formal scaling domain.

The constrained classical equation is

\begin{equation}
[A^\nu,[A_\nu,A_\mu]]
=
2\lambda_QA_\mu.
\end{equation}

Under \(A_\mu\mapsto bA_\mu\), the left-hand side scales as \(b^3\), while \(A_\mu\) on the right-hand side scales as \(b\). Therefore,

\begin{equation}
\lambda_Q(u;s)
=
b(s)^2
\lambda_Q(u;9).
\end{equation}

The coupling-free quartic quantity and the bosonic action scale as

\begin{equation}
S_{B,0}(u;s)
=
b(s)^4
S_{B,0}(u;9),
\end{equation}

and

\begin{equation}
S_B(u;s)
=
b(s)^4
S_B(u;9).
\end{equation}

These relations are homogeneous scaling identities and do not modify the coupling convention of Section 2.

The constrained physical Hessian likewise scales as

\begin{equation}
H_{\rm phys}(u;s)
=
b(s)^2
H_{\rm phys}(u;9).
\end{equation}

It follows that the zero mode tangent to the physical modulus remains exactly zero under this continuation. Every transverse hard eigenvalue is multiplied by the same positive factor \(b(s)^2\), so the numbers of positive and negative hard directions are unchanged.

The tangent along the physical modulus scales as

\begin{equation}
\partial_uA_\mu(u;s)
=
b(s)\,
\partial_uA_\mu(u;9).
\end{equation}

For the auxiliary positive-definite coefficient-space metric introduced in Appendix D, this implies

\begin{equation}
G_{uu}(u;s)
=
b(s)^2
G_{uu}(u;9).
\end{equation}

Accordingly, the corresponding coefficient-space proper-length element satisfies

\begin{equation}
d\ell_s
=
b(s)d\ell_9
=
b(s)
\sqrt{G_{uu}(u;9)}\,du.
\end{equation}

This proper length is an auxiliary coefficient-space quantity used in the collective-coordinate construction. It is not a Lorentzian spacetime distance. The form above also makes clear that \(u\) itself is not assumed to be an exact proper-length coordinate.

As a numerical spot check, we also evaluated several nearby radial slices around \(s=9\). In these checks, the homogeneously rescaled backgrounds continued to satisfy the constrained equations to the same floating-point accuracy, the physical nullity remained one, and no additional hard zero mode appeared. This supports regularity of the homogeneous continuation in a neighborhood of \(s=9\), but it is not used to establish a certified continuous interval or a global radial range for the branch.

\subsection{Radial factorization of the one-loop density}\label{radial-factorization-of-the-one-loop-density}

The quadratic operators introduced in Appendix C scale homogeneously along the continuation:

\begin{equation}
H_{\rm hard}(u;s)
=
b(s)^2
H_{\rm hard}(u;9),
\end{equation}

\begin{equation}
\Delta_{\rm FP}(u;s)
=
b(s)^2
\Delta_{\rm FP}(u;9),
\end{equation}

and

\begin{equation}
\mathcal M_F(u;s)
=
b(s)
\mathcal M_F(u;9).
\end{equation}

The \(169\)-dimensional hard bosonic Gaussian therefore scales as

\begin{equation}
|\det H_{\rm hard}|^{-1/2}
\sim
b^{-169}.
\end{equation}

The \(24\)-dimensional Faddeev--Popov determinant scales as

\begin{equation}
\det\Delta_{\rm FP}
\sim
b^{48},
\end{equation}

while the Pfaffian of the \(384\times384\) antisymmetric Majorana--Weyl operator has degree \(192\) and therefore scales as

\begin{equation}
\operatorname{Pf}\mathcal M_F
\sim
b^{192}.
\end{equation}

Consequently, the root-referenced intrinsic fermionic scalar is unchanged across the homogeneous slices,

\begin{equation}
\Delta\mathcal P_F(u;s)
=
\Delta\mathcal P_F(u;9).
\end{equation}

Equivalently, the fixed-section determinant diagnostic satisfies

\begin{equation}
\Gamma_{1,\mathrm{sec}}^{\det}(u;s)
-
\Gamma_{1,\mathrm{sec}}^{\det}(u;9)
=
169\log b(s)
-
48\log b(s)
-
192\log b(s),
\end{equation}

or

\begin{equation}
\Gamma_{1,\mathrm{sec}}^{\det}(u;s)
-
\Gamma_{1,\mathrm{sec}}^{\det}(u;9)
=
-71\log b(s).
\end{equation}

This radial shift is independent of \(u\). It refers to the same transported Lorentz section and auxiliary hard-basis convention; it does not promote \(\Gamma_{1,\mathrm{sec}}^{\det}\) to a quotient scalar.

We next consider the collective-coordinate factor in the explicit-FP convention of Appendix D. Under the same homogeneous rescaling,

\begin{equation}
\sqrt{G_{uu}}
\sim b,
\end{equation}

the \(24\)-dimensional color-gauge orbit factor scales as

\begin{equation}
J_G\sim b^{24},
\end{equation}

and the projected Lorentz-frame orbit factor scales as

\begin{equation}
J_L\sim b^{45}.
\end{equation}

Here \(45\) is the actual projected Lorentz-frame tangent rank on the analyzed branch.

The fixed-\(Q_\rho\) restriction gives

\begin{equation}
J_Q\sim b.
\end{equation}

The geometric part of the collective-coordinate measure therefore scales as

\begin{equation}
\sqrt{G_{uu}}
\frac{J_GJ_L}{J_Q}
\sim
b^{1+24+45-1}
=
b^{69}.
\end{equation}

Since Appendix D keeps the Faddeev--Popov determinant explicitly in the fluctuation factor, \(J_{\rm mod}\) contains its inverse. Therefore,

\begin{equation}
J_{\rm mod}(u;s)
=
b(s)^{21}
J_{\rm mod}(u;9).
\end{equation}

Combining all contributions gives

\begin{equation}
21+48+192-169=92.
\end{equation}

Thus, the local one-loop density magnitude in the fixed continuation coordinate \(u\) obeys the exact factorization

\begin{equation}
\mathcal W_{\rm 1loop}(u,s)
=
b(s)^{92}
\mathcal W_{\rm 1loop}(u,9).
\end{equation}

Equivalently, the exponent may be written directly from the geometric factors as

\begin{equation}
92
=
1_{\sqrt{G_{uu}}}
+
24_{J_G}
+
45_{J_L}
-
1_{J_Q}
+
192_{\rm Pf}
-
169_{\rm hard\ boson}.
\end{equation}

This gives the complete origin of the radial power \(92\).

The density one-form, rather than its coordinate coefficient, is the reparametrization-invariant object. Since

\begin{equation}
d\ell_s
=
b(s)d\ell_9,
\end{equation}

the density coefficient with respect to the auxiliary proper-length element scales instead as

\begin{equation}
\mathcal W_\ell(u;s)
=
b(s)^{91}
\mathcal W_\ell(u;9).
\end{equation}

The \(b^{92}\) and \(b^{91}\) statements therefore refer to different density coefficients and are not in conflict.

For each radial slice, define

\begin{equation}
W_{\rm rel}(u;s)
=
\frac{
\mathcal W_{\rm 1loop}(u,s)
}{
\mathcal W_{\rm 1loop}(0,s)
}.
\end{equation}

The common radial factor cancels exactly:

\begin{equation}
W_{\rm rel}(u;s)
=
W_{\rm rel}(u;9).
\end{equation}

Thus, within this homogeneous continuation, the relative local one-loop profile along the physical modulus is not an artifact of the particular normalization slice \(Q_\rho=9\). This is an exact local factorization statement; it does not extend the numerically verified physical branch beyond the neighborhood analyzed in F.1.

\subsection{Formal reduced radial model and global Lorentzian scope}\label{reduced-radial-integral-and-global-lorentzian-scope}

The exact real-positive radial factorization also motivates a conditional one-variable reduction. The density magnitude constructed in Appendix D is defined on real backgrounds and contains absolute values of determinant/Pfaffian factors together with positive coefficient-space Jacobians. By itself, that real magnitude does not define a holomorphic function of a complex radial variable. Any complex-\(r\) ray used below must therefore be understood as a formal reduced model, or equivalently as a conditional analytic continuation after compatible branches for the determinant/Pfaffian and Jacobian factors have been chosen. No such choice is inferred from the real magnitude alone.

Within that qualification, the factorization gives the following conditional statement about integration in the radial direction.

Suppose the same radial integration prescription is applied for every value of \(u\) on this factorized homogeneous branch. Since the classical action scales quartically,

\begin{equation}
S_B(u;s)
=
b(s)^4S_B(u;9),
\end{equation}

we obtain

\begin{equation}
\int ds\,
\mathcal W_{\rm 1loop}(u,s)
e^{iS_B(u;s)}
=
\mathcal W_{\rm 1loop}(u,9)
\int ds\,
b(s)^{92}
e^{ib(s)^4S_{B,0}(9)/g^2}.
\end{equation}

The classical action \(S_B(u;9)\) is constant along the physical modulus, so the second factor is common to all \(u\). Therefore, under a common radial prescription, radial integration contributes a common multiplicative factor and does not remove the relative \(u\)-profile.

This is a conditional statement about the factorized homogeneous branch. It is not a computation of the full Lorentzian IIB matrix integral.

For the reduced one-variable problem, set

\begin{equation}
r=b(s).
\end{equation}

With

\begin{equation}
s+Q_{\rm ref}
=
D_0r^2,
\qquad
D_0=9+Q_{\rm ref}=15+2\sqrt3,
\end{equation}

we have

\begin{equation}
ds
=
2D_0r\,dr.
\end{equation}

On the reference slice, the coupling-independent quartic coefficient is

\begin{equation}
S_{B,0}(9)
:=
S_{B,0}[A(u;9)]
=
S_{B,0}[A(0;9)]
\simeq
25.5903936986,
\end{equation}

and the action convention of Section 2 gives

\begin{equation}
S_B(9)
=
\frac{S_{B,0}(9)}{g^2}.
\end{equation}

Thus the numerical value above refers to \(S_{B,0}(9)\) rather than to a choice of the coupling \(g\). For the standard real positive coupling, the positivity of \(S_{B,0}(9)/g^2\) is a numerical property of this branch and does not imply that the Lorentzian bosonic action is positive definite in general.

The reduced radial factor then takes the form

\begin{equation}
2D_0
\int dr\,
r^{93}
e^{iS_{B,0}(9)r^4/g^2}.
\end{equation}

This radial integral is conceptually different from the local Lorentzian Gaussian of Appendix D. There, an infinitesimal positive damping factor is introduced in the finite-dimensional real fluctuation coefficient space around each real background. Here, by contrast, \(r\) is a single homogeneous radial variable obtained after the local factorization.

For the formal complex continuation of the reduced variable, and after a compatible branch choice for the factors entering the reduced measure, the numerical property

\begin{equation}
S_{B,0}(9)/g^2>0
\end{equation}

implies that the phase factor

\begin{equation}
e^{iS_{B,0}(9)r^4/g^2}
\end{equation}

decays along the complex-\(r\) directions

\begin{equation}
\arg r
\in
\left\{
\frac{\pi}{8},
\frac{5\pi}{8},
\frac{9\pi}{8},
\frac{13\pi}{8}
\right\}.
\end{equation}

For example, define the inner reduced ray integral along

\begin{equation}
r=e^{i\pi/8}t,
\qquad
0\le t<\infty,
\end{equation}

by

\begin{equation}
I_{\rm ray}
:=
\int_0^{e^{i\pi/8}\infty}
dr\,
r^{93}
e^{iS_{B,0}(9)r^4/g^2}.
\end{equation}

The overall radial Jacobian factor \(2D_0\) is not included in this definition.

Under this formal branch choice and the standard damped/ray prescription, the reduced integral is

\begin{equation}
I_{\rm ray}
=
\frac14
e^{-i\pi/4}
\Gamma_{\rm Euler}
\left(
\frac{47}{2}
\right)
\left[\frac{S_{B,0}(9)}{g^2}\right]^{-47/2}.
\end{equation}

Here \(\Gamma_{\rm Euler}\) denotes the Euler gamma function, to distinguish it from \(\Gamma_{1,\mathrm{sec}}^{\det}\). This is only the mathematical value of the formally continued reduced contour integral under the stated branch choice. It is not a holomorphic continuation supplied by the real density magnitude itself, and it is not, by itself, the physical contribution of this branch to the IIB matrix integral.

The reduced measure behaves as

\begin{equation}
r^{93}dr
\end{equation}

near \(r=0\) and is integrable there. However, \(r=0\) corresponds to

\begin{equation}
A_\mu=0.
\end{equation}

At this point, the original quotient description becomes singular, and in particular the fixed-\(Q_\rho\) coarea factor satisfies

\begin{equation}
J_Q=0.
\end{equation}

Therefore, regularity of the reduced one-variable endpoint does not imply uniform validity of the one-loop approximation in the full matrix configuration space.

The three levels of the Lorentzian problem must therefore be kept distinct:

\begin{longtable}[]{@{}
  >{\raggedright\arraybackslash}p{(\columnwidth - 2\tabcolsep) * \real{0.5000}}
  >{\raggedright\arraybackslash}p{(\columnwidth - 2\tabcolsep) * \real{0.5000}}@{}}
\toprule\noalign{}
\begin{minipage}[b]{\linewidth}\raggedright
object
\end{minipage} & \begin{minipage}[b]{\linewidth}\raggedright
status
\end{minipage} \\
\midrule\noalign{}
\endhead
\bottomrule\noalign{}
\endlastfoot
local finite-dimensional \(i\epsilon\) Gaussian around each real background & defined in Appendix D \\
reduced one-variable radial integral after local factorization & analytically characterized here \\
full matrix-space Lorentzian integration cycle & not determined \\
thimble decomposition and intersection coefficients for the physical cycle & not determined \\
\end{longtable}

In particular, the reduced steepest-decay ray does not determine the physical full matrix-space integration cycle or its intersection coefficients. These remain global questions outside the analysis of this paper.

\subsection{Algebraic special point and the physical modulus}\label{algebraic-special-point-and-the-physical-modulus}

The historical algebraic ansatz used during the construction of the backgrounds contains the special parameter values

\begin{equation}
Y=\sqrt3,
\qquad
X=3+2\sqrt3.
\end{equation}

The scalar symbols \(X\) and \(Y\) in this subsection refer only to this historical ansatz. They are distinct from the sector tensors

\begin{equation}
\mathsf X_{\mu\nu},
\qquad
\mathsf Y_{\mu\nu}
\end{equation}

used in Appendices A and E.

For completeness, we record the exact historical ansatz. Let \(E_{ab}\) denote the \(5\times5\) matrix unit, and define

\begin{equation}
T_{ab}
=
i(E_{ab}-E_{ba}),
\qquad
\operatorname{Tr}(T_{ab}^2)=2.
\end{equation}

Let the three historical orbit scales be denoted by

\begin{equation}
t,\qquad a_{\rm sc},\qquad v.
\end{equation}

The historical scalar parameters are

\begin{equation}
X
=
\frac{t^2}{a_{\rm sc}^2},
\qquad
Y
=
\frac{v^2}{a_{\rm sc}^2},
\qquad
q_{\rm sc}
=
a_{\rm sc}^2.
\end{equation}

The symbol \(a_{\rm sc}\) is used here so that the manuscript-wide symbol \(u\) remains reserved for the physical-modulus coordinate.

In the unit convention

\begin{equation}
q_{\rm sc}=1,
\end{equation}

define

\begin{equation}
\alpha
=
\sqrt{3+2\sqrt3},
\qquad
\beta
=
3^{1/4}.
\end{equation}

The exact special configuration is then

\begin{equation}
\begin{aligned}
(A_0,\ldots,A_9)
=
\bigl(
&\alpha T_{12},
T_{13},
T_{14},
T_{15},
T_{23},
T_{24},
T_{25},\\
&\beta T_{34},
\beta T_{35},
\beta T_{45}
\bigr).
\end{aligned}
\end{equation}

These equations completely specify the ten traceless Hermitian \(5\times5\) matrices of the historical special configuration.

At this special point,

\begin{equation}
Q_\rho=0,
\qquad
S_B=0.
\end{equation}

However, direct evaluation of the classical equation gives three different component blocks:

\begin{equation}
[A^\nu,[A_\nu,A_0]]
=
6A_0,
\end{equation}

\begin{equation}
[A^\nu,[A_\nu,A_M]]
=
0,
\qquad
M=1,\ldots,6,
\end{equation}

and

\begin{equation}
[A^\nu,[A_\nu,A_M]]
=
(4+2\sqrt3)A_M,
\qquad
M=7,8,9.
\end{equation}

Since all ten matrices are nonzero, no single value of \(\lambda_Q\) can satisfy

\begin{equation}
[A^\nu,[A_\nu,A_\mu]]
=
2\lambda_QA_\mu
\end{equation}

for all components simultaneously. The special configuration is therefore not a constrained stationary solution.

The off-shell quadratic operators at this point have

\begin{equation}
\text{ghost nullity}=6,
\qquad
\text{bosonic quadratic nullity}=21.
\end{equation}

These zero eigenvalues belong to quadratic operators constructed around an off-shell configuration. They are not the physical constrained-Hessian zero mode found on the stationary branch of Section 3.

The nonlinear behavior found in this off-shell analysis also differs from the physical modulus. In the off-shell quadratic zero directions, the cubic term vanishes, while generic directions are lifted at quartic order. A five-dimensional special locus remains in this off-shell analysis. None of these facts turns the special point into the origin of the one-dimensional physical modulus.

One can also rescale the special configuration so that it has

\begin{equation}
Q_\rho=9.
\end{equation}

The required factor is

\begin{equation}
b_*^2
=
\frac{13-3\sqrt3}{4}.
\end{equation}

After this rescaling, the normalized constrained residual is

\begin{equation}
0.326.
\end{equation}

Thus, homogeneous rescaling does not convert the historical special configuration into a constrained root on the \(Q_\rho=9\) slice.

The physical modulus studied in this paper has a different origin. It is obtained from an actual zero mode of the constrained physical Hessian at a stationary solution and is then continued nonlinearly while satisfying the full constrained classical equations.

The historical special point should therefore be regarded only as an algebraic construction seed. It is not on the representative physical branch, it is not a fixed-\(Q_\rho=9\) constrained saddle, and it is not the origin of the physical modulus.

\subsection{Stationary-solution landscape}\label{stationary-solution-landscape}

The physical modulus analyzed in Sections 3--6 is not the only stationary structure found in the bounded numerical search.

A bounded multi-start search found several inequivalent stationary classes, and more than one class admitted a verified local one-dimensional continuation. The search was not designed as an exhaustive classification of the fixed-\(Q_\rho\) stationary solution space.

The branch used in this paper is one representative modulus-bearing class. It is fixed by the classical stationary and continuation analysis; the one-loop and higher-\(L\) responses are evaluated only afterwards and are not used as branch-selection criteria. We therefore make no claim of uniqueness, statistical representativeness, or dynamical priority for this branch.

A comparison of the physical contributions of the different stationary classes would require their fluctuation factors together with a prescription for the full Lorentzian integration cycle. Those ingredients are not determined by the present analysis.

\end{document}